\documentclass[a4paper,11pt]{article}
\pdfoutput=1 % if your are submitting a pdflatex (i.e. if you have
\usepackage{jcappub} % for details on the use of the package, please
\usepackage[T1]{fontenc} % if needed
\usepackage[dvipsnames]{xcolor}

\title{Fast Simulation of Cosmological Neutral Hydrogen based on the Halo Model}

\author[a,1]{Pascal Hitz,\note{Corresponding author.}}
\author[a]{Pascale Berner,}
\author[a]{Devin Crichton,}
\author[a,b]{John Hennig,}
\author[a]{and Alexandre Refregier}

\affiliation[a]{Institute for Particle Physics and Astrophysics, ETH Zurich, 8092 Zurich, Switzerland}
\affiliation[b]{Scientific IT Services, ETH Zurich, 8092 Zurich, Switzerland}

\emailAdd{hitzpa@phys.ethz.ch}
\emailAdd{bernerp@phys.ethz.ch}
\emailAdd{dcrichton@phys.ethz.ch}
\emailAdd{john.hennig@id.ethz.ch}
\emailAdd{alexandre.refregier@phys.ethz.ch}

\abstract{Cosmological neutral hydrogen (HI) surveys provide a promising tomographic probe of the post-reionization era and of the standard model of cosmology. Simulations of this signal are crucial for maximizing the utility of these surveys. We present a fast method for simulating the cosmological distribution of HI based on a halo model approach. Employing the approximate \texttt{PINOCCHIO} code, we generate the past light cone of dark matter halos. Subsequently, the halos are populated with HI according to a HI-halo mass relation. The nature of \qty{21}{\cm} intensity mapping demands large-volume simulations with a high halo mass resolution. To fulfill both requirements, we simulate a past light cone for declinations between \qty{-15}{\degree} and \qty{-35}{\degree} in the frequency range from 700 to \qty{800}{\mega\hertz}, matching HIRAX, the Hydrogen Intensity and Real-time Analysis eXperiment. We run \texttt{PINOCCHIO} for a \qty{1}{h^{-3} Gpc^3} box with $6700^3$ simulation particles. With this configuration, halos with masses as low as $M_\text{min} = \qty{4.3e9}{{M}_{\odot}}$ are simulated, resulting in the recovery of more than \qty{97}{\%} of the expected HI density. From the dark matter and HI past light cone, maps with a width of \qty{5}{\mega\hertz} are created. To validate the simulation, we have implemented and present here an analytical dark matter and HI halo model in \texttt{PyCosmo}, a Python package tailored for theoretical cosmological predictions. We perform extensive comparisons between analytical predictions and the simulation for the mass function, mass density, power spectrum, and angular power spectrum for dark matter and HI. We find close agreement in the mass function and mass densities, with discrepancies within a few percent. For the three-dimensional power spectra and angular power spectra, we observe an agreement better than \qty{10}{\%}. This approach is broadly applicable for forecasting and forward-modeling of dedicated intensity mapping experiments, such as HIRAX, as well as cosmological surveys with the SKA. Additionally, it is particularly well suited to be extended for cross-correlation studies. The \texttt{PyCosmo} halo model extension and simulation datasets are made publicly available.}

\begin{document}
\maketitle

\flushbottom

\section{Introduction}
\label{sec:Introduction}
The measurement of the recent accelerated expansion of the universe confronts us with a fundamental problem for our understanding of its nature. Current cosmological models invoke an unknown dark energy, which is now the dominant energy component in the universe. To shed light on dark energy, its characteristic properties need to be empirically determined, in particular the equation of state. In order to achieve this, the expansion history of the universe must be constrained in tomographic analyses, starting from the epoch when dark energy began to influence the cosmic acceleration, around a redshift of $z \sim 2$. Baryonic Acoustic Oscillations (BAO) provide a unique standard ruler which can be used to measure the expansion history. Measuring the observed scale of the BAO in the matter power spectrum in the radial and transverse directions allows to measure the Hubble parameter and the comoving angular-diameter distance. These in turn provide information about the expansion history \cite{Eisenstein_1998, Seo&Eisenstein_2003}. \\
The BAO have been well-studied in galaxy surveys before \cite{DESI_III_2024, DESI_IV_2024, DESI_VI_2024, Alam_2021, Alam_2017}, but there exists another promising technique with several advantages: \qty{21}{\cm} intensity mapping \cite{Battye_2004, Peterson_2006, Chang_2008, Wyithe_2007}. After the end of reionization, neutral hydrogen (HI) is expected to be a good tracer of the underlying matter distribution \cite{Liu_2020}. Even in its ground state HI can be observed through the emission of radiation due to the hyperfine transition between the state in which the spins of the electron and proton are parallel aligned to the state in which they are anti-aligned. This transition produces a photon of about \qty{21}{\cm} wavelength corresponding to a rest frequency of $f_{21} = \qty{1420.406}{\mega\hertz}$. The fact that this is a forbidden transition is compensated by the vast amount of HI and is an advantage because the probability that the photon is absorbed again after emission is very small. Since the expansion of the universe causes the emitted photons to be redshifted, the spectral line of this transition is observed at a lower frequency $f_\text{obs}$. The relation between the observed frequency and redshift is given by:
\begin{equation}
    z = \frac{f_{21}}{f_\text{obs}}-1.
    \label{eq:redshift-frequency relation}
\end{equation}
Consequently, individual galaxies do not have to be resolved to infer their redshift, as is the case with galaxy surveys. Instead, mapping the emission of unresolved HI as a function of frequency immediately provides three-dimensional information about its distribution. While post-reionization \qty{21}{\cm} intensity mapping can target the same science as galaxy surveys, it has the advantage of observing much larger volumes, i.e. wide fields in a broad redshift range, with a lower angular resolution sufficient for cosmological scales, in a shorter time. It therefore has the potential to constrain cosmological parameters to very high precision. \\
There are several instruments aiming to perform \qty{21}{\cm} intensity mapping both in single-dish mode (e.g. SKAO-MID \cite{SKAO_2015, SKAO_2018}, MeerKAT \cite{MeerKAT_2016}, BINGO \cite{BINGO_2019}, FAST \cite{FAST_2011}, GBT \cite{GBT_2009}, and Parkes \cite{Parkes_1996}) and in interferometric mode (e.g. CHIME \cite{CHIME_2014}, CHORD \cite{CHORD_2019}, HERA \cite{HERA_2017}, PUMA \cite{PUMA_2019}, Tianlai \cite{Tianlai_2015}, and HIRAX \cite{HIRAX_2016, HIRAX_2022}). HIRAX, which stands for the Hydrogen Intensity and Real-time Analysis eXperiment, is a radio interferometer currently in development. It will be located in the Karoo desert in South Africa covering the southern sky from the celestial equator to a declination of \qty{-60}{\degree}. Eventually, HIRAX will consist of 1024 six meter dishes arranged in a redundant configuration and operating in drift-scan mode, i.e. scanning the sky for constant but adjustable declinations through the natural rotation of the Earth. HIRAX will be sensitive to the frequency range of 400 to \qty{800}{\mega\hertz} and therefore probe in the redshift range from 0.78 to 2.55. Although HIRAX is still under construction and therefore observational data is not yet available, simulations of the \qty{21}{\cm} signal in the form of tomographic maps can be used to test the instrument and observation modeling as well as the analysis pipeline. This work focuses on simulating post-reionization HI signal maps and verifying their quality. \\
On one end, statistical simulations, such as lognormal simulations \cite{Lognormal_1991, FLASK_2016, cora_2020, CoLoRe_2022}, provide a computationally efficient option to generate many realisations of a known statistical distribution. However, their accuracy is limited by our understanding of the input distribution, which is often what we seek to improve through simulations. On the other end, large \textit{N}-body or hydrodynamical simulations can model complex non-linear processes that are analytically intractable, but their high computational costs prevent us from running a large number of realisations. Lagrangian Perturbation Theory (LPT) based methods aim to bridge the gap between the two approaches, providing accurate realisations of the matter distribution up to mildly non-linear scales while accepting some inaccuracies on very small scales. One approach is to use LPT to compute the evolution of the dark matter density field and subsequently applying a global bias model to link HI to this on the field level \cite{CoLoRe_2022}. However, in the post-reionization era, most HI is expected to reside within dense regions, i.e. in dark matter halos, where it can self-shield from ionizing radiation. Although intensity mapping instruments typically have low angular resolution, they have a high frequency resolution in the radial direction. Consequently, shot noise, arising from the discrete nature of the HI distribution, becomes increasingly significant. To capture this more realistic description of HI and to account for the discreteness of the sources, we employ the \texttt{PINOCCHIO} code \cite{PINOCCHIO_2002a, PINOCCHIO_2002b, PINOCCHIO_2002c, PINOCCHIO_2013, PINOCCHIO_2017a, PINOCCHIO_2017b}. This code uses LPT to efficiently simulate the past light cone of dark matter halos in an approximate but quicker way than full \textit{N}-body simulations. The HI is then subsequently assigned to these halos using a HI-halo mass relation that has been fitted to observations. Due to its approximate nature, the small-scale clustering and halo positions of \texttt{PINOCCHIO} are probably not accurate enough for future simulation-based inference on real observational data. However, to generate forecasts for our intended applications, this level of accuracy is sufficient. \\
This method has been previously applied in cosmological HI signal simulations for the SKAO HI intensity blind foreground subtraction challenge described in \cite{Spinelli_2021}. There, a full-sky past light cone was generated and a HI-halo mass relation was used, which was calibrated using a semi-analytical model \cite{Spinelli_2020}. In this work, we limit the construction of the light cone to the region of the sky covered by the HIRAX telescope, allowing us to simulate a larger number of particles with \texttt{PINOCCHIO}. This is essential for accurately capturing most of the cosmological HI. \\
In future work, these maps will serve to test the HIRAX pipeline used to reconstruct the HI emission signal from our simulated signal. Furthermore, the fact that the simulations are based on consistent halo catalogues opens up the possibility for upcoming cross-correlation studies. This paper is dedicated to the development and running of a HI simulation pipeline and the validation of the simulated signal by comparison with analytical predictions. For the latter, we have implemented an extension including a theoretical dark matter and HI halo model in \texttt{PyCosmo} \footnote{PyCosmo webpage: \url{https://cosmology.ethz.ch/research/software-lab/PyCosmo.html}.} \cite{PyCosmo_Refregier, PyCosmo_Tarsitano, PyCosmo_Moser}, a Python-based framework for theoretical cosmological predictions. In particular, predictions of the mass function, mass densities and non-linear power spectra are compared to the simulations. As part of this work, we present the halo model and the specific implementations that are made publicly available in \texttt{PyCosmo 2.2.0}. \\
This paper is organised as follows: In section \ref{sec:Analytical Methods}, all details of the implemented dark matter and HI halo model are described. Section \ref{sec:Simulation Methods} proceeds with the procedure of simulating the past light cone of dark matter and HI halos and the map generation step. Section \ref{sec:Results} presents the resulting dark matter and HI temperature maps and compares the simulated mass function, mass densities, power spectra and angular power spectra to predictions within the framework of the dark matter and HI halo model. Finally, section \ref{sec:Conclusions} serves as a summary of this paper and provides an outlook to future work. \\
Throughout this paper, a flat $\Lambda$CDM cosmology with the cosmological parameters from the Planck 2018 results \cite{Planck_2018} is assumed: $\Omega_\text{m} = 0.3153$, $\Omega_{\Lambda} = 0.6847$, $\Omega_\text{b} = 0.0493$, $h = 0.6736$, $\sigma_8 = 0.8111$, $n_\text{s} = 0.9649$, and $Y_\text{p} = 0.24$.

\section{Analytical Methods}
\label{sec:Analytical Methods}
Our simulations are based on the distribution of dark matter halos. Therefore, we make use of the halo model formalism \cite{Peacock_Smith_2000, Seljak_2000, Cooray_Sheth_2002} to make theoretical predictions for comparison with the simulated data. The fundamental assumption of this formalism is that all matter in the universe is arranged in halos of different sizes and masses. Given that HI is expected to trace these halos, we can apply a halo model analogous to that used for dark matter to HI as well. At the heart of every halo model are three ingredients: the halo mass function, the halo bias, and the halo profile. Once these are defined, the formalism can be used to calculate the mean mass densities, the non-linear three-dimensional power spectrum and, consequently, the angular power spectrum. \\
Given the various implementations and conventions, and for the convenience of the reader, we give a detailed review of the halo model in the following sections as we have implemented it in \texttt{PyCosmo}. 

\subsection{Dark Matter Halo Model}
\label{sec:Dark Matter Halo Model}
\subsubsection{Ingredients}
\label{sec:Ingredients}
The mass function, which gives the number of dark matter halos of mass \textit{M} per unit volume and unit mass at a given redshift \textit{z}, is given as:
\begin{equation}
    \frac{dn(M,z)}{dM} = \frac{\overline{\rho}_\text{m,0}}{M^2} \nu f(\nu) \bigg| \frac{d\ln\nu}{d\ln M} \bigg|.
    \label{eq:mass function}
\end{equation}
The peak height is defined as $\nu(M,z) = \delta_\text{c}(\Omega_\text{m}(z))/\sigma(M,z)$, with $\delta_\text{c}$ the critical overdensity needed for spherical collapse of a halo and $\sigma^2(M,z)$ being the mass variance of the density field smoothed on the scale $R = (3 M / (4 \pi \overline{\rho}_\text{m,0}))^{1/3}$. In the spherical collapse model regions with mean overdensity larger than the critical overdensity are assumed to have collapsed into a virialized object. This criterion can either be evaluated at the present time or at the time of collapse. In the first case, $\delta_\text{c}$ must be linearly extrapolated to today's value by dividing by the linear growth factor. In the later case, the density field must be considered at this redshift and the mass variance is linearly extrapolated to the specific redshift. We decide to use the second convention and make the mass variance redshift dependent by multiplying it by the linear growth factor. For the multiplicity function $f(\nu)$, we provide four different options in \texttt{PyCosmo}: Press \& Schechter \cite{Press_Schechter_1974}, Sheth \& Tormen \cite{Sheth_Tormen_1999}, Tinker et al. \cite{Tinker_2010}, and Watson et al. \cite{Watson_2013}. For the mass function of Tinker et al., the critical overdensity is set to $\delta_\text{c}=\frac{3}{20}(12 \pi)^{2/3} \approx 1.686$, the value found in spherical collapse in a flat matter dominated universe $\Omega_\text{m}=1$, and we have implemented the mass function with the parameters found for $\Delta_\text{v}=200$ in \cite{Tinker_2010}. \footnote{See equation \ref{eq:virial radius} for the formal definition of $\Delta_\text{v}$.} For the other mass functions, we allow cosmology dependent expressions \cite{Nakamura_Suto_1997, Mo_White_2010}. For a flat $\Lambda$CDM model, $\Omega_\Lambda \neq 0$ and $\Omega_\Lambda + \Omega_\text{m} = 1$:
\begin{equation}
    \delta_\text{c}(\Omega_\text{m}(z)) = \frac{3}{20} (12 \pi)^{2/3} \Omega_\text{m}(z)^{0.0055},
\label{eq:delta_c_lambda}
\end{equation}
and for $\Omega_\Lambda = 0$ and $\Omega_\text{m} \neq 1$:
\begin{equation}
    \delta_\text{c}(\Omega_\text{m}(z)) = \frac{3}{20} (12 \pi)^{2/3} \Omega_\text{m}(z)^{0.0185}.
\label{eq:delta_c_nolambda}
\end{equation}
Ultimately, we analyse a simulation that is in comoving coordinates and so we use today's mass density $\overline{\rho}_\text{m,0}$ in equation (\ref{eq:mass function}). This means that the halo mass function, as it is defined here, captures only the growth of structure due to gravitational attraction as a function of time, but it factors out the expansion of the universe. \footnote{If the expansion of the universe should be included, one has to use $\overline{\rho}_\text{m}(z) = \overline{\rho}_\text{m,0}(1+z)^3$ in the definition of the halo mass function.} \\
Using the halo mass function, the mean dark matter density can be calculated as an integral over the dark matter mass. To predict the density in the simulations at a specific redshift, the lower and upper bound of the integral are given by the minimal and maximal simulated halo mass at that time $\mathbf{M}_\text{cut}(z) = [M_\text{min}(z), M_\text{max}(z)]$:
\begin{equation}
    \overline{\rho}_\text{m,0}\big(z;\mathbf{M}_\text{cut}(z)\big) = \int\displaylimits_{M_\text{min}(z)}^{M_\text{max}(z)} M \, \frac{dn(M,z)}{dM} \, dM.
    \label{eq:matter rhobar}
\end{equation}
This expression is again in comoving coordinates and the redshift dependency only describes the hierarchical growth of halos, but it does not include the expansion of the universe. \\
The fundamental assumption of the halo model, that all dark matter is arranged in halos, $\overline{\rho}_\text{m,0} = \overline{\rho}_\text{m,0}(z;M_\text{min}=0,M_\text{max}=\infty)$, results in a normalisation condition for the mass function or multiplicity function, respectively:
\begin{equation}
    1 \equiv \frac{1}{\overline{\rho}_\text{m,0}} \int\displaylimits_0^\infty M \, \frac{dn(M,z)}{dM} \, dM = \int\displaylimits_0^\infty f(\nu) \, d\nu.
    \label{eq:normalisation1}
\end{equation} \\
In \texttt{PyCosmo}, we also provide four different linear halo biases to describe how halos cluster on large scales relative to the underlying matter field: Mo and White \cite{Mo_White_1996}, Sheth \& Tormen \cite{Sheth_Tormen_1999}, Sheth, Mo \& Tormen \cite{Sheth_Mo_Tormen_2001}, and Tinker et al. \cite{Tinker_2010}. As for the mass function of Tinker et al., we have also implemented the bias parameters for $\Delta_\text{v} = 200$ found in \cite{Tinker_2010}. \\
The spatial distribution of dark matter within individual halos is modeled using a Navarro-Frenk-White (NFW) profile \cite{NFW_1997}:
\begin{equation}
    \rho(r,M) = \frac{\rho_\text{s}}{\Big(\frac{r}{r_\text{s}}\Big) \Big(1 + \frac{r}{r_\text{s}}\Big)^2}.
    \label{eq:NFW profile}
\end{equation}
The profile is parameterised by the scale radius $r_\text{s}$ and the normalisation $\rho_\text{s}$. The scale radius is usually defined by the halo mass dependent concentration parameter, which is the ratio between the virial radius of a halo with mass \textit{M}, $r_\text{vir}(M)$, and the scale radius: $c(M) = r_\text{vir}(M) / r_\text{s}$. For the concentration parameter, we make use of the fitting function from \cite{Bullock_2001}  and the virial radius is given as:
\begin{equation}
    r_\text{vir}(M) = \bigg(\frac{3 M}{4 \pi \overline{\rho}_\text{m,0} \Delta_\text{v}} \bigg)^{1/3}.
    \label{eq:virial radius}
\end{equation}
As for the critical overdensity for spherical collapse, we also allow for a cosmology dependent expression for $\Delta_\text{v}$ \cite{Bryan_Norman_1998}. \footnote{Since $r_\text{vir}$ in equation (\ref{eq:virial radius}) is defined with respect to the matter density rather than the critical density, we have an additional $1/\Omega_\text{m}(z)$ compared to \cite{Bryan_Norman_1998}.} For a flat $\Lambda$CDM cosmology, $\Omega_\Lambda \neq 0$ and $\Omega_\Lambda + \Omega_\text{m} = 1$:
\begin{equation}
    \Delta_\text{v} = \frac{18 \pi^2 + 82 d(z) - 39 d(z)^2}{\Omega_\text{m}(z)},
    \label{eq:delta_v_lambda}
\end{equation}
and for $\Omega_\Lambda = 0$ and $\Omega_\text{m} \neq 1$:
\begin{equation}
    \Delta_\text{v} = \frac{18 \pi^2 + 60 d(z) - 32 d(z)^2}{\Omega_\text{m}(z)},
    \label{eq:delta_v_nolambda}
\end{equation}
with $d(z) = \Omega_\text{m}(z)-1$. Both expressions simplify to $\Delta_\text{v} = 18 \pi^2 \approx 178$ for $\Omega_\text{m} = 1$. The normalisation $\rho_\text{s}$ results from the requirement that integrating the profile over a sphere with a radius equal to the virial radius returns the total halo mass \textit{M}. This leads to the following expression:
\begin{equation}
    \rho_\text{s} = \frac{\Delta_\text{v}\overline{\rho}_\text{m,0}}{3} \frac{c(M)^3}{\ln(1+c(M)) - c(M)/(1+c(M))}.
    \label{eq:NFW normalization}
\end{equation}

\subsubsection{Dark Matter Power Spectrum}
\label{sec:DM Power Spectrum}
After defining the mass function, the halo bias, and the halo profile, the non-linear dark matter power spectrum can be calculated. Therefore, we adopt the convention given by the Fourier transform of the dark matter overdensity field, $\delta_\text{m}(\vec{x},z) = (\rho_\text{m}(\vec{x},z) - \overline{\rho}_\text{m}(z)) / \overline{\rho}_\text{m}(z)$: \footnote{$\delta^{(3)}_\text{D}$ denotes the three-dimensional Dirac delta function.}
\begin{equation}
    \big<\tilde{\delta}_\text{m}(\vec{k},z)\tilde{\delta}^{*}_\text{m}(\vec{k'},z) \big> = (2\pi)^3 \delta^{(3)}_\text{D}(\vec{k} - \vec{k'}) P_\text{DM}(k,z).
    \label{eq:def_pk}
\end{equation}
In the halo model formalism, the power spectrum can be divided into a one- and two-halo term, $P_\text{DM,1h}$ and $P_\text{DM,2h}$, respectively:
\begin{equation}
    P_\text{DM}\big(k,z;\mathbf{M}_\text{cut}(z)\big) = P_\text{DM,1h}\big(k,z;\mathbf{M}_\text{cut}(z)\big) + P_\text{DM,2h}\big(k,z;\mathbf{M}_\text{cut}(z)\big).
    \label{eq:total power spectrum}
\end{equation}
While the one-halo term describes the contributions that come from correlations within a single halo, the two-halo term describes the correlations between different halos. 
As with the mean dark matter density, the two terms can be written as integrals over the halo mass \textit{M}. To enable a direct comparison with the simulations, the lower and upper limits are again given by the smallest and largest simulated halos, $M_\text{min}(z)$ and $M_\text{max}(z)$:
\begin{align}
    &P_\text{DM,1h}(k,z) 
    = \frac{1}{\overline{\rho}_\text{m,0}^2\big(z;\mathbf{M}_\text{cut}(z)\big)} \int\displaylimits_{M_\text{min}(z)}^{M_\text{max}(z)} \, \frac{dn(M,z)}{dM} \, M^2 \, |\tilde{u}(k|M)|^2 \, dM 
    \\[2pt]
    &P_\text{DM,2h}(k,z) 
    = \frac{P_\text{lin}(k,z)}{\overline{\rho}_\text{m,0}^2\big(z;\mathbf{M}_\text{cut}(z)\big)} \biggl[ \int\displaylimits_{M_\text{min}(z)}^{M_\text{max}(z)} \, \frac{dn(M,z)}{dM} \, M \, b(M,z) \, \tilde{u}(k|M) \, dM \biggr]^2.
    \label{eq:One- and Two-halo term dark matter}
\end{align}
These expressions include the linear matter power spectrum $P_\text{lin}(k,z)$, for which we use the fit of Eisenstein and Hu \cite{Eisenstein_Hu_1999}, and the Fourier transform of the normalised spherically symmetric halo profile truncated at the virial radius $\tilde{u}(k|M)$. If we use the NFW profile from equation (\ref{eq:NFW profile}), we get \cite{Cooray_Sheth_2002}: 
\begin{equation}
    \begin{split}
    \tilde{u}(k|M) &= \frac{4 \pi}{M} \int\displaylimits_0^{r_\text{vir}(M)} \, \rho(r,M) \, \frac{\sin(kr)}{kr} \, r^2 dr
    \\
    &=  \begin{aligned}[t]
            \frac{4 \pi \rho_\text{s} r_\text{s}^3}{M} \bigg(&\sin(k r_\text{s}) \Big[\text{Si}([1+c(M)]k r_\text{s}) - \text{Si}(k r_\text{s}) \Big] - \frac{\sin(c(M) k r_\text{s})}{(1+c(M)) k r_\text{s}} \\
            &+ \cos(k r_\text{s}) \Big[\text{Ci}([1+c(M)] k r_\text{s}) - \text{Ci}(k r_\text{s}) \Big] \bigg).
        \end{aligned}
    \end{split}
    \label{eq:fourier transform ukm}
\end{equation}
In this expression, we make use of the sine and cosine integrals:
\begin{equation}
    \text{Ci}(x) = - \int\displaylimits_x^\infty \frac{\cos t}{t} dt \quad\mathrm{and}\quad \text{Si}(x) = \int\displaylimits_0^x \frac{\sin t}{t} dt.
    \label{eq:sine and cosine integral}
\end{equation}
However, since in our simulations the halos are considered as point sources and their profile is therefore a Dirac delta function, $\tilde{u}(k|M)=1$ in our case. It is worth noting that regardless of whether it is assumed that the halos have a profile or not, on very large scales (i.e. for small \text{k}) $\tilde{u}(k|M) \rightarrow 1$. If it is also considered that on large scales the two-halo term has to approach $P_\text{lin}(k,z)$, we find a second normalisation condition similar to equation (\ref{eq:normalisation1}):
\begin{equation}
    1 \equiv \frac{1}{\overline{\rho}_\text{m,0}} \int\displaylimits_0^\infty \, \frac{dn(M,z)}{dM} \, M \, b(M,z) \, dM = \int\displaylimits_0^\infty f(\nu) \, b(\nu) \, d\nu.
    \label{eq:normalisation2}
\end{equation}

\subsubsection{Dark Matter Angular Power Spectrum}
\label{sec:DM Angular Power Spectrum}
The final products of our simulation pipeline are sky maps of the projected dark matter distribution with a given width in the radial direction. The key statistic that we are interested in is thus the angular power spectrum. It is a three-dimensional integral in Fourier space and along the radial comoving distance $\chi(z)$ of the power spectrum. For a flat cosmology, it takes the following form: 
\begin{equation}
    \begin{split}
    C_\text{$\ell$,DM} = &\frac{2}{\pi} \int  k^2 dk \int\displaylimits^{\infty}_0  d\chi W(\chi) j_{\ell}(k \chi) \int\displaylimits^{\infty}_0 d\chi' W(\chi') j_{\ell}(k \chi')\\
    &\times \sqrt{P_\text{DM}\big(k,z(\chi); \mathbf{M}_\text{cut}(z(\chi))\big) \cdot P_\text{DM}\big(k,z(\chi'); \mathbf{M}_\text{cut}(z(\chi'))\big)}.
    \end{split}
    \label{eq:Cl_dm}
\end{equation}
The power spectrum is weighted by window functions $W(\chi(z))$ and spherical Bessel functions $j_{\ell}(k\chi)$. It should be noted that for non-flat cosmologies, the spherical Bessel functions are replaced by hyper-spherical Bessel functions. In our case, $W(\chi(z))$ is a top hat function in the redshift interval of a map and is therefore only non-zero in a very limited redshift or $\chi(z)$ range. Furthermore, the function is normalised as:
\begin{equation}
    \int\displaylimits_0^\infty W(\chi) d\chi = 1.
    \label{eq:window function normalisation}
\end{equation}
Since equation (\ref{eq:Cl_dm}) involves three nested integrals, it is common to use the Limber approximation \cite{Limber_1953} to accelerate the computation:
\begin{equation}
    \begin{split}
        C_\text{$\ell$,DM}
        &\simeq \int d \chi \frac{W^2(\chi)}{r^2(\chi)} P_\text{DM}\biggl(k = \frac{\ell + 1/2}{r(\chi)}, z(\chi); \mathbf{M}_\text{cut}(z(\chi)) \biggr) \\
        &= \int dz \frac{c}{H(z)} \frac{W^2(z)}{r^2(z)} P_\text{DM}\biggl(k = \frac{\ell + 1/2}{r(z)}, z; \mathbf{M}_\text{cut}(z) \biggr),
    \end{split}
\label{eq:Cl_dm_limber}
\end{equation}
with the angular diameter distance $r(\chi)$ and the Hubble parameter $H(z)$. The Limber approximation simplifies further in the case of a very narrow window function. Under the assumption that the power spectrum and the angular diameter distance remain constant over the limited integration range, they can be evaluated at a fixed $\chi^*$ and factored out of the integral. The normalisation of the window function then ensures that no integration is required:
\begin{equation}
    C_\text{$\ell$,DM} \simeq \frac{W(\chi^*)}{r^2(\chi^*)} P_\text{DM}\biggl(k = \frac{\ell + 1/2}{r(\chi^*)}, z(\chi^*); \mathbf{M}_\text{cut}(z(\chi^*)) \biggr).
\label{eq:Cl_dm_limber_narrow_window}
\end{equation}
When $r(\chi)$ is used in the expression of the Limber approximation, it is generally valid for flat and non-flat cosmologies. For a flat cosmology, it is straight forward to derive equation (\ref{eq:Cl_dm_limber}) from equation (\ref{eq:Cl_dm}) \cite{Dodelson_Modern_Cosmology_2020}: By using that the product of two spherical Bessel functions for high $\ell$ has a sharp peak around $k\chi \approx k\chi' \approx \sqrt{\ell(\ell+1)} \approx \ell+1/2$, $P_\text{DM}$ can be taken out of the integration over \textit{k} as long as it is approximately constant over the peak. Subsequently using the orthogonal property of spherical Bessel functions \footnote{$\frac{2}{\pi} \int k^2 dk j_{\ell}(k \chi) j_{\ell}(k \chi') = \delta_\text{D}^{(1)}(\chi - \chi') / \chi^2$, where $\delta_\text{D}^{(1)}$ denotes the one-dimensional Dirac delta function.} and remembering that $r(\chi) = \chi$ for the flat case finally results in equation (\ref{eq:Cl_dm_limber}). The result for non-flat cosmologies can be recovered by generalising the calculations found in e.g. \cite{Dodelson_Modern_Cosmology_2003, Kaiser_1992} for the flat case by using $r(\chi)$ in the definition of the comoving position $\Vec{x} \approx (r(\chi) \theta_1, r(\chi) \theta_2, \chi)$ instead of just $\chi$. \\
However, the Limber approximation tends to diverge for narrow window functions, as is the case for our maps. We will see this in our results for the angular power spectrum (see section \ref{sec:Angular Power Spectrum}) when we compare it to the spectrum measured from our maps. \\
Since performing the complete integration in equation (\ref{eq:Cl_dm}) is numerically challenging, particularly for very narrow redshift bins, we opted to use a Monte-Carlo integration. In future work, we aim to explore more advanced methods, e.g. \cite{Angpow_2017, FFTLog_2018}, to perform the integration. Note that this computation is not implemented in \texttt{PyCosmo}, but we make use of its halo model power spectrum to perform the computation. However, the Limber approximation is fully integrated into the package and functional.

\subsection{HI Halo Model}
\label{sec:HI Halo Model}
Similarly to the dark matter halo model, one can also describe a HI halo model, which only needs minor changes compared to the former. Since in the post-reionization era HI is assumed to reside mainly in dark matter halos, we model the HI mass distribution by assigning HI mass to the dark matter halos through a $M_\text{HI}$-\textit{M} relation. Two of the before mentioned ingredients for a complete halo model can be inherited from the dark matter halo model: the halo mass function and the halo bias. We only introduce a new HI halo profile which replaces the dark matter halo profile from the last section.

\subsubsection{HI-Halo Mass Relation}
\label{sec:HI-Halo Mass Relation}
To assign the right amount of HI to the dark matter halos, we use a HI-halo mass relation that has been constrained by HI data in \cite{Padmanabhan_2017}. As constraints, they utilized data from the small-scale clustering, column density and mass function of HI galaxies at low redshift, intensity mapping measurements at intermediate redshifts and UV/optical observations of Damped Lyman Alpha systems at higher redshifts. The amount of HI mass in a dark matter halo of mass \textit{M} as a function of redshift is given as:
\begin{equation}
    M_\text{HI}(M,z) = \alpha f_\text{H,c} M \biggl (\frac{M}{10^{11} \, h^{-1} \text{M}_{\odot}} \biggr)^{\beta} \exp \biggl[- \Bigl( \frac{v_\text{c,0}}{v_\text{c}(M,z)} \Bigr)^3 \biggr].
    \label{eq:MHI-M relation}
\end{equation}
The overall normalisation $\alpha$ stands for the mass fraction of HI in halos relative to the cosmic fraction $f_\text{H,c} = (1-Y_\text{p})\Omega_\text{b} / \Omega_\text{m}$. The logarithmic slope of the relation is described by $\beta$. The minimum virial velocity, $v_\text{c,0}$, corresponds to a minimum virial halo mass of halos that are still able to host HI. In \cite{Padmanabhan_2017}, they find the best-fit values $\alpha = 0.09$, $\beta = -0.58$, and $\log v_\text{c,0} = 1.56$ for these parameters. Although they assumed a slightly different cosmology \footnote{$\Omega_\text{m} = 0.281$, $\Omega_\Lambda = 0.719$, $h = 0.71$, $\Omega_\text{b} = 0.0462$, $\sigma_8 = 0.8$, $n_\text{s} = 0.96$, and $Y_\text{p} = 0.24$}, we also use these values in this work. We use this relation with the best-fit parameter values in both the theoretical analysis and the simulation although they could be varied to account for uncertainties on their values. For the virial velocity as a function of halo mass, we use \cite{Barnes&Haehnelt_2014}:
\begin{equation}
    v_\text{c}(M,z) = \SI{96.6}{\km \, \s^{-1}} \biggl( \frac{\Delta_\text{v} \Omega_\text{m} h^2}{24.4} \biggr)^{1/6} \biggl( \frac{1 + z}{3.3} \biggr)^{1/2} \biggl( \frac{M}{10^{11} \, \text{M}_{\odot}} \biggr)^{1/3}.
    \label{eq:v_cM}
\end{equation}
The relationship between the HI mass and the halo mass describes an exponential cut-off at low halo masses, which is due to the fact that the halos are too small to self-shield HI from ionization and thus host HI. The redshift dependence of the virial velocity leads to less and less HI being hosted in small halos as time evolves. \\ 
With this prescription for the assignment of HI to the halos, the mean comoving HI density at a given redshift for given minimum and maximum dark matter halo masses can again be written as an integral over the dark matter mass. The halo mass function only has to be weighted with the HI mass instead of dark matter mass:
\begin{equation}
    \overline{\rho}_\text{HI,0}\big(z;\mathbf{M}_\text{cut}(z)\big) = \int\displaylimits_{M_\text{min}(z)}^{M_\text{max}(z)} \, M_\text{HI}(M,z) \, \frac{dn(M,z)}{dM} \, dM.
    \label{eq:HI rhobar}
\end{equation}

\subsubsection{HI Halo Profile}
\label{sec:HI Halo Profile}
To characterise the HI halo profile, we also follow \cite{Padmanabhan_2017}. In observations and simulations \cite{Obreschkow_2009, Wang_2014} it has been found that the HI distribution can be modelled by a spherically symmetric exponential profile in the radial direction:
\begin{equation}
    \rho(r,M) = \rho_0 \exp(-r/r_\text{s,HI}).
    \label{eq:density profile}
\end{equation}
Here $r_\text{s,HI} = r_\text{vir}(M)/c_\text{HI}(M,z)$ is now the scale radius of HI defined through the concentration parameter $c_\text{HI}(M,z)$:
\begin{equation}
    c_\text{HI}(M,z) = c_\text{HI,0} \biggl( \frac{M}{10^{11} \, \text{M}_{\odot}} \biggr) ^{-0.109} \frac{4}{(1+z)^{\gamma}},
    \label{eq:concentration parameter}
\end{equation}
with the best-fit values found in \cite{Padmanabhan_2017} for the two parameters: $c_\text{HI,0} = 28.65$ and $\gamma = 1.45$.
The normalisation constant of the profile is approximated to be $\rho_0 = M_\text{HI}(M,z) / (8 \pi r_\text{s}^3)$ such that the HI mass inside the virial radius integrates to $M_\text{HI}(M,z)$.

\subsubsection{HI Power Spectrum}
\label{sec:HI Power Spectrum}
Analogous to the dark matter power spectrum, we adopt the convention for the non-linear HI power spectrum given by the Fourier transform of the HI mass overdensity field, $\delta_\text{HI}(\vec{x},z) = (\rho_\text{HI}(\vec{x},z) - \overline{\rho}_\text{HI}(z)) / \overline{\rho}_\text{HI}(z)$:
\begin{equation}
    \big<\tilde{\delta}_\text{HI}(\vec{k},z)\tilde{\delta}^{*}_\text{HI}(\vec{k'},z) \big> = (2\pi)^3 \delta^{(3)}_\text{D}(\vec{k} - \vec{k'}) P_\text{HI}(k,z).
    \label{eq:def_pk_HI}
\end{equation}
It can be expressed again as the sum of a one- and two-halo term, which are given as:
\begin{align}
    &P_\text{HI,1h}(k,z) 
    = \frac{1}{\overline{\rho}_\text{HI,0}^2\big(z;\mathbf{M}_\text{cut}(z)\big)} \int\displaylimits_{M_\text{min}(z)}^{M_\text{max}(z)} \, \frac{dn(M,z)}{dM} \, M_\text{HI}^2(M,z) \, |\tilde{u}_\text{HI}(k|M)|^2 \, dM 
    \\[2pt]
    &P_\text{HI,2h}(k,z) 
    = \frac{P_\text{lin}(k,z)}{\overline{\rho}_\text{HI,0}^2\big(z;\mathbf{M}_\text{cut}(z)\big)} \biggl[ \int\displaylimits_{M_\text{min}(z)}^{M_\text{max}(z)} \, \frac{dn(M,z)}{dM} \, M_\text{HI}(M,z) \, b(M,z) \, \tilde{u}_\text{HI}(k|M) \, dM \biggr]^2.
    \label{eq:HI one- and two-halo term}
\end{align}
The Fourier transform of the normalised HI halo profile truncated at the virial radius is given by:
\begin{equation}
    \tilde{u}_\text{HI}(k|M) = \frac{4\pi}{M_\text{HI}(M,z)} \int\displaylimits_0^{r_\text{vir}(M)} \, \rho(r,M) \frac{\sin(kr)}{kr}r^2 \, dr.
    \label{fourier transform}
\end{equation}
Substituting the exponential profile (\ref{eq:density profile}) into this expression, we obtain:
\begin{equation}
    \tilde{u}_\text{HI}(k|M) = \frac{4 \pi \rho_0 r_\text{s,HI}^3 u_1(k|M)}{M_\text{HI}(M,z)},
    \label{uHI}
\end{equation}
with
\begin{equation}
    u_1(k|M) = \frac{2}{(1+k^2r_\text{s,HI}^2)^2}.
    \label{u1}
\end{equation}
In comparison to our halo-based HI simulations, we again consider point sources and therefore use $\tilde{u}_\text{HI}(k|M) = 1$.
\subsubsection{HI Angular Power Spectrum}
\label{sec:HI Angular Power Spectrum}
The full expression of the HI angular power spectrum and its Limber approximation are again obtained by integrating the three-dimensional HI power spectrum. The full expression for a flat cosmology is given by:
\begin{equation}
    \begin{split}
    C_\text{$\ell$,HI} = &\frac{2}{\pi} \int  k^2 dk \int\displaylimits^{\infty}_0 d\chi W(\chi) j_{\ell}(k \chi) \int\displaylimits^{\infty}_0 d\chi' W(\chi') j_{\ell}(k \chi')\\
    &\times \sqrt{P_\text{HI}\big(k,z(\chi); \mathbf{M}_\text{cut}(z(\chi))\big) \cdot P_\text{HI}\big(k,z(\chi'); \mathbf{M}_\text{cut}(z(\chi'))\big)}.
    \label{eq:Cl_HI}
    \end{split}
\end{equation}
And the Limber approximation takes the form:
\begin{equation}
    \begin{split}
        C_\text{$\ell$,HI}
        &\simeq \int d \chi \frac{W^2(\chi)}{r^2(\chi)} P_\text{HI}\biggl(k = \frac{\ell + 1/2}{r(\chi)}, z(\chi); \mathbf{M}_\text{cut}(z(\chi)) \biggr) \\
        &= \int dz \frac{c}{H(z)} \frac{W^2(z)}{r^2(z)} P_\text{HI}\biggl(k = \frac{\ell + 1/2}{r(z)}, z; \mathbf{M}_\text{cut}(z) \biggr),
    \end{split}
\label{eq:Cl_HI_limber}
\end{equation}
which further simplifies again when assuming a very narrow window function:
\begin{equation}
    C_\text{$\ell$,HI} \simeq \frac{W(\chi^*)}{r^2(\chi^*)} P_\text{HI}\biggl(k = \frac{\ell + 1/2}{r(\chi^*)}, z(\chi^*); \mathbf{M}_\text{cut}(z(\chi^*)) \biggr).
\label{eq:Cl_HI_limber_narrow_window}
\end{equation}
As already for dark matter, equation (\ref{eq:Cl_HI}) is performed with a Monte-Carlo integration and is not implemented in \texttt{PyCosmo}, but the Limber approximation is.

\section{Simulation Methods}
\label{sec:Simulation Methods}
Having discussed the theoretical halo model in detail, we turn our attention in the following sections to the procedure for generating \qty{21}{\cm} emission maps based on a halo model approach. First, we describe the simulation of the dark matter halo distribution, i.e. the simulation of a past light cone halo catalogue, which is subsequently populated with HI using equation (\ref{eq:MHI-M relation}). We also explain how maps are generated from the past light cone.

\subsection{Simulation of Dark Matter and HI Halos}
\label{sec:Simulation of Dark Matter Halos and HI Halos}
\begin{figure}
    \centering
    \includegraphics[width=0.7\textwidth]{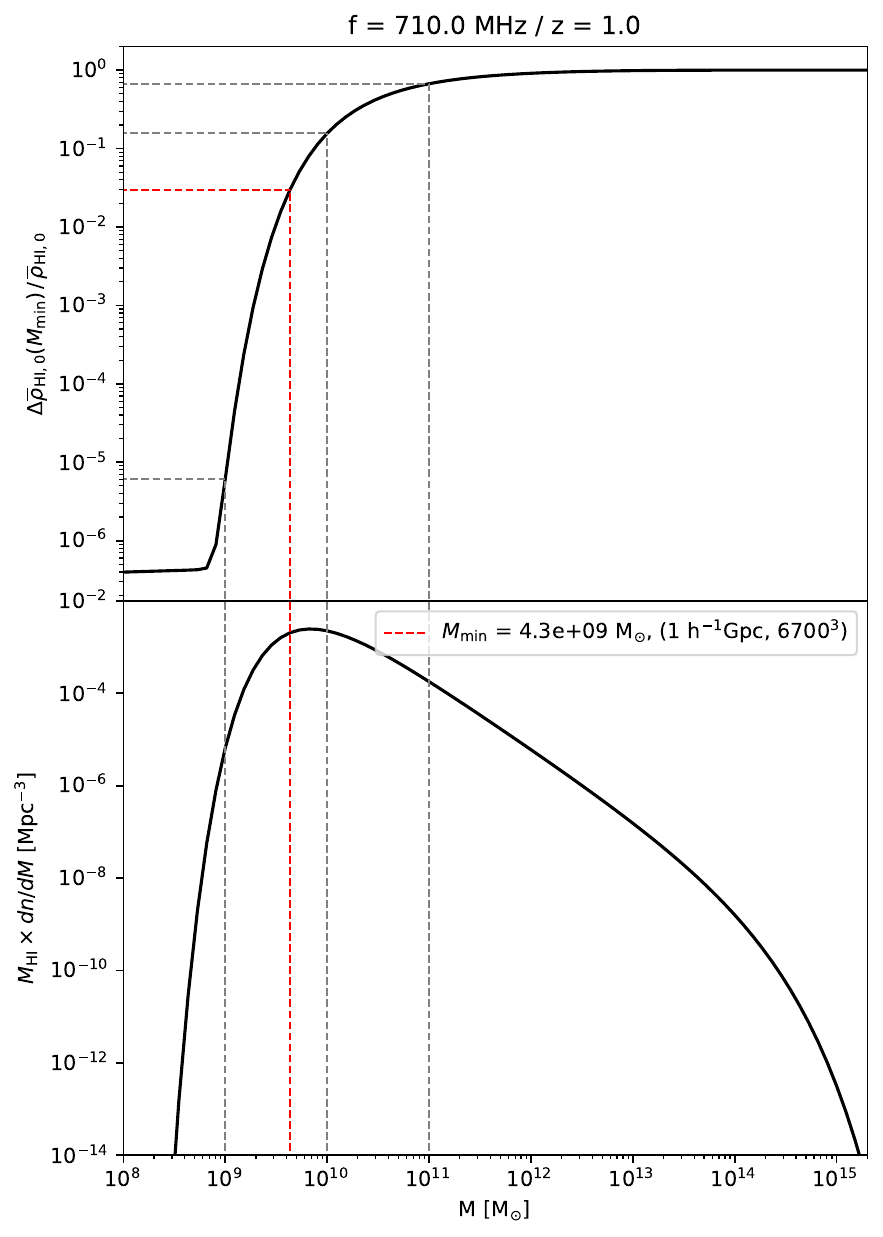}
    \caption{\textbf{Upper panel:} Loss in the mean HI density as a function of the minimum simulated dark matter halo mass $M_\text{min}$ relative to the total HI density. In the calculation of the density the mass function of Watson \cite{Watson_2013} is assumed. The red dashed line indicates the loss for the simulated minimum halo mass achieved for our simulation with a box size of \qty{1}{h^{-1} Gpc} and $6700^3$ simulation particles per box. The grey dashed lines are shown for better visualisation.
    \textbf{Lower panel:} Integrand of the mean HI density as given in equation (\ref{eq:HI rhobar}).}
    \label{fig:hi_loss}
\end{figure}

As already mentioned in section \ref{sec:Introduction}, we try to seek an intermediate solution between realistic but computationally intensive \textit{N}-body simulations and computationally efficient but simplified statistical simulations. Therefore, we use the \texttt{PINOCCHIO} code \footnote{In this work, we use version 4.1.3 available on \url{https://github.com/pigimonaco/Pinocchio}} (PINpointing Orbit Crossing Collapsed HIerarchical Objects) \cite{PINOCCHIO_2002a, PINOCCHIO_2002b, PINOCCHIO_2002c, PINOCCHIO_2013, PINOCCHIO_2017a, PINOCCHIO_2017b}. It has been shown to be much faster while still recovering clustering properties at the \qty{10}{\%}-level when compared to \textit{N}-body. The massive speed-up is due to the usage of Lagrangian Perturbation Theory together with the Extended Press \& Schechter formalism \cite{Bond_1991} for non-spherical collapse to calculate the masses, positions, velocities, and merger histories of dark matter halos. Furthermore, \texttt{PINOCCHIO} generates the past light cone halo catalogue with continuous time sampling on-the-fly. \\
The goal of intensity mapping is to observe large cosmological volumes, e.g. a comoving volume of approximately \qty{90}{h^{-3} Gpc^3} with HIRAX, at low angular resolution rather than resolving individual objects. Therefore, our interest lies in the simulation of such vast volumes. However, although there is more HI in large dark matter halos, as we have discussed in section \ref{sec:HI-Halo Mass Relation}, there are many more small halos than large ones, making their contribution to the total HI content not negligible. Thus, not only large volumes are required, but also a sufficiently good mass resolution, which would be hard to achieve with conventional full \textit{N}-body simulations. Figure \ref{fig:hi_loss} quantifies how much HI is lost at redshift $z=1$ for a given mass resolution. The upper panel shows the relative difference in the mean HI density, as described in equation (\ref{eq:HI rhobar}), for different lower mass cuts, compared to the total density when no cut is applied. The bottom panel shows the integrand of $\overline{\rho}_\text{HI,0}$. \footnote{For this figure the mass function of Watson is assumed.} The decrease of the integrand at the high mass end results from the rarity of high mass halos described by the mass function. The exponential decrease at the low mass end results from the inability of these halos to self-shield HI from ionization. However, this exponential decrease facilitates to simulate a significant part of HI with a limited resolution in the first place: As long as halo masses just below the peak of the integrand are simulated, most of the HI is captured. As described in section \ref{sec:HI-Halo Mass Relation}, this cut-off shifts to lower masses with increasing redshift, which requires a better mass resolution of the simulation.
 \\
Since, in future work, we are interested in applying our generated \qty{21}{\cm} maps to the instrument simulation and analysis pipeline of HIRAX, we choose our light cone geometry according to the survey configuration. Given that memory requirements are a limiting factor for the simulation, a good balance must be achieved between the volume of the past light cone and the mass resolution. HIRAX will operate in drift-scan mode, observing strips of constant declination. Therefore, instead of running a full-sky simulation, we can confine the simulation to the volume covered by HIRAX, allowing computational resources to be allocated toward maximizing the mass resolution. We decide to simulate the past light cone to cover declinations between \qty{-15}{\degree} and \qty{-35}{\degree} and a frequency range from 700 to \qty{800}{\mega\hertz}, that means for redshifts from 0.77 to 1.03. Pushing \texttt{PINOCCHIO} to its limits for this light cone geometry, we manage to run a simulation for a box size of \qty{1}{h^{-1} Gpc} and $6700^3$ simulation particles per box, choosing that the smallest identified halos consist of at least ten particles. To cover the cosmological volume, 40 boxes are replicated and stacked together with periodic boundary conditions. Although the full light cone has a width of \qty{0.46}{h^{-1} Gpc}, which is about half the box size, the geometry of the light cone requires two box replications in the radial direction. In the transverse direction, the circumference of the light cone is between 10.4 and \qty{14.2}{h^{-1} Gpc} at $z=1.03$ and between 8.3 and \qty{11.4}{h^{-1} Gpc} at $z=0.77$. This requires approximately between 20 and 24 box replications, depending on the distance from the central observer. Box replications will introduce artefacts in the clustering signal at scales corresponding to the box size and larger. With the given configuration, the smallest simulated halos have masses of \qty{4.3e9}{M_{\odot}}, which is indicated by the red dashed line in figure \ref{fig:hi_loss}. At $z=1$, the simulation is therefore missing around \qty{3}{\%} of the total HI. The missing fraction shrinks to around \qty{2}{\%} at $z=0.77$. \\
The displacements of the particles are calculated with third-order LPT, while the grouping of the collapsed particles into halos, a step called fragmentation, is performed with second-order LPT. For fragmentation, the simulation box is divided into sub-volumes. Each sub-volume is assigned to a separate MPI task. Since the different tasks do not communicate with each other, the calculations need to be extended to a boundary layer that overlaps between the sub-volumes. The thickness of the boundary layer is defined as the Lagrangian radius of the largest halo expected at the last redshift multiplied by a boundary layer factor set in the \texttt{PINOCCHIO} parameter file. As recommended in the documentation, \footnote{\url{https://github.com/pigimonaco/Pinocchio/blob/master/DOCUMENTATION}} we set this parameter to 1, which leads to a significant overhead in memory requirements. The simulation is run on 2400 nodes with 12 cores each on the supercomputer Piz Daint (Cray XC40/50) \cite{Piz_Daint}, hosted by the Swiss National Supercomputing Centre (CSCS). It requires a total of $\sim$\qty{150}{TB} of memory and $\sim$40'000 CPU hours.

\subsection{Map Generation}
\label{sec:Map Generation}
Since the great advantage of intensity mapping is the direct mapping between the frequency of the observation and redshift, we generate thin maps from the past light cone by projecting equally spaced bins in frequency to resolve the structure along the line of sight. To generate pixelated maps on a sphere, the software package \texttt{HEALPix} \footnote{\url{http://healpix.sourceforge.net}} \cite{HEALPix}, in particular the Python package \texttt{healpy} \cite{healpy}, is used. The \textit{NSIDE} parameter, which defines the number of pixels on the map, is set to 512. This corresponds to an angular resolution of about \qty{6.9}{\arcminute}, which is just below the maximum resolution of \qty{12}{\arcminute} of the initially planned HIRAX 256-element array \cite{HIRAX_2022}. The frequency resolution of HIRAX will be \qty{0.39}{\mega \hertz} \cite{HIRAX_2022}. However, as illustrated in appendix \ref{sec:Poisson Noise Contribution to the Angular Power Spectrum}, we found that for maps with such a narrow bandwidth, Poisson noise (see also section \ref{sec:Angular Power Spectrum}) dominates the cosmological signal for a wide range of scales in the angular power spectrum due to the small number of discrete halos per pixel. Since we are mainly interested here in the simulated cosmological signal and whether it matches theoretical expectations, we choose a coarser bandwidth of \qty{5}{\mega \hertz}, resulting in 20 separate maps. In this work, we deliberately use cosmological redshifts rather than observed redshifts, as the latter would introduce additional complexities in comparing our results to theoretical predictions. We plan to incorporate observed redshifts and their implications in future work. A thorough analysis of PINOCCHIO's accuracy in simulating the velocity field will be essential for that. Nevertheless, prior studies have demonstrated good agreement with an \textit{N}-body simulation in the redshift space power spectrum when PINOCCHIO employs second-order LPT for halo construction and third-order LPT for halo displacements \cite{PINOCCHIO_2017a}. To create dark matter- or HI-mass maps, the pixels are weighted by the masses of the halos falling into them. However, the key observable of a radio interferometer is the brightness temperature. To convert the HI mass maps into HI brightness temperature maps, we use \cite{Bull_2015}:
\begin{equation}
    T_\text{b}(\vec{x},z) = \frac{3 h c^3 A_{10}}{32 \pi k_\text{B} f_{21}^2 m_\text{p}} \frac{(1+z)^2}{H(z)} \rho_\text{HI}(\vec{x},z),
    \label{eq:brightness temperature}
\end{equation}
where $h$ is the Planck constant, $c$ the speed of light, $A_{10} \approx \qty{2.869e-15}{s^{-1}}$ \cite{Wilson_2009} the Einstein coefficient for spontaneous emission, $k_\text{B}$ the Boltzmann constant, and $m_\text{p}$ the proton mass. To calculate $\rho_\text{HI}$, the HI mass maps are divided by the volume of a voxel. \footnote{The volume is calculated by dividing the volume of a \qty{5}{\mega\hertz} thick shell by the number of pixels.} The redshift dependent factor $(1+z)^2 / H(z)$ is calculated by averaging the redshifts of all halos that fall within a pixel. \footnote{We have found that the central redshift of each bin would also be sufficient for this expression, since the bins are small. However, this would not work for wider bins in redshift.} \\
It is important to note that when comparing the power spectra of the simulation with the theoretical predictions outlined in sections \ref{sec:HI Power Spectrum} and \ref{sec:HI Angular Power Spectrum}, converting the HI mass into brightness temperature is not necessary. As long as overdensities are calculated initially, the spectra in equations (\ref{eq:def_pk_HI}) and (\ref{eq:Cl_HI}) are computed. This is because the brightness temperature is proportional to the HI mass density:
\begin{equation}
    \delta T_\text{b}(\vec{x},z) = \frac{T_\text{b}(\vec{x},z) - \overline{T}_\text{b}(z)}{\overline{T}_\text{b}(z)} = \frac{\rho_\text{HI}(\vec{x},z) - \overline{\rho}_\text{HI}(z)}{\overline{\rho}_\text{HI}(z)} = \delta_\text{HI}(\vec{x},z).
\end{equation}

\section{Results}
\label{sec:Results}
In the following sections, we will compare the statistics of the simulations with the theoretical predictions based on the halo model. This serves as a check to ensure that the simulated signal agrees with theoretical expectations before applying the simulations to instrument-related analyses. The compared statistics are the mass function, the dark matter and HI mass densities as a function of redshift, the three-dimensional dark matter and HI power spectra, and the corresponding angular power spectra.
\begin{figure}
    \centering
    \includegraphics[width=1.0\textwidth]{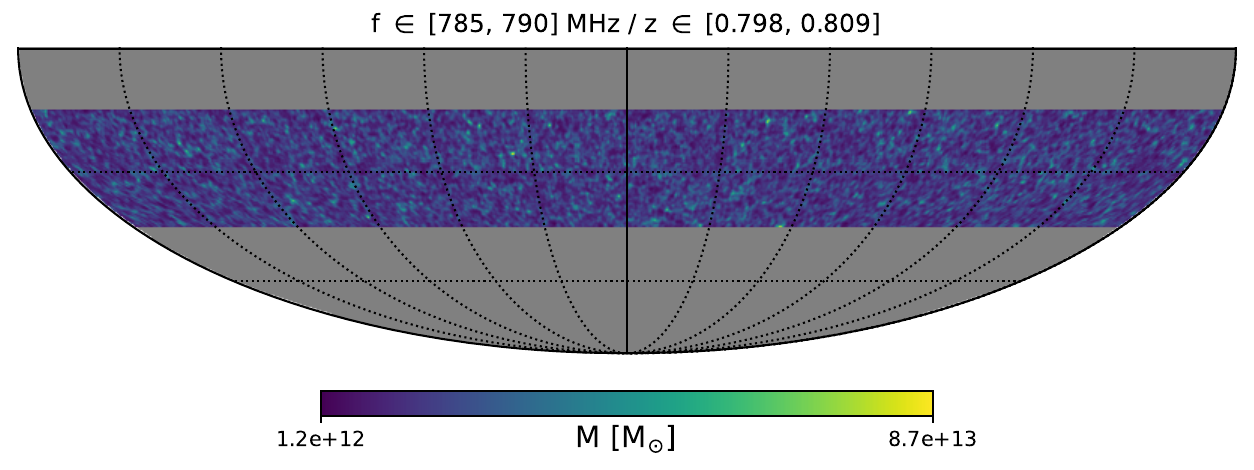}
    \caption{Dark matter map for the frequency bin from 785 to \qty{790}{\mega\hertz}, smoothed with a Gaussian kernel with $\sigma = \qty{0.005}{\radian} \approx \qty{17.2}{\arcminute}$, generated from a halo light cone produced with \texttt{PINOCCHIO}.}
    \label{fig:dm_mass_map}
\end{figure}
\begin{figure}
    \centering
    \includegraphics[width=1.0\textwidth]{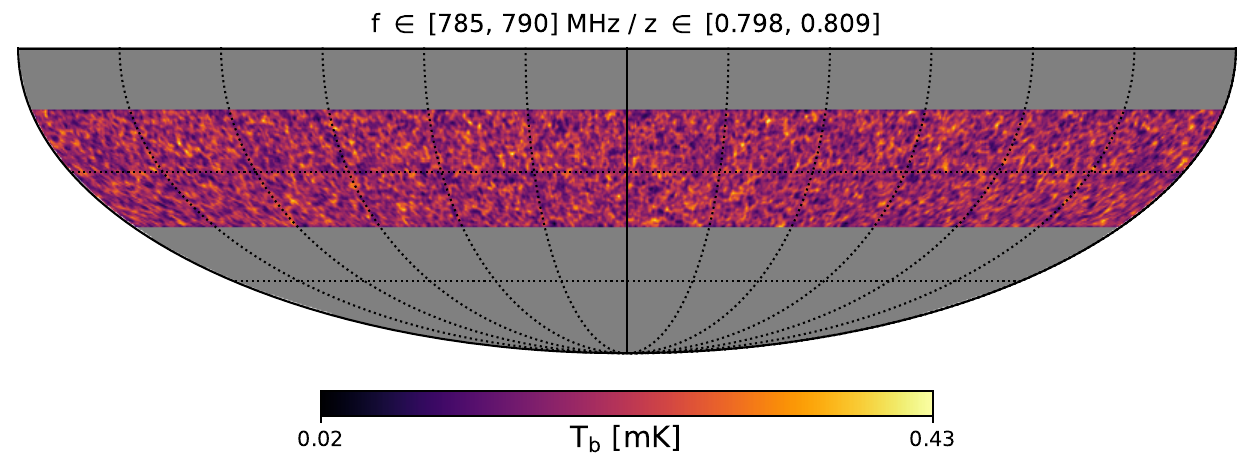}
    \caption{HI brightness temperature map for the frequency bin from 785 to \qty{790}{\mega\hertz}, smoothed with a Gaussian kernel with $\sigma = \qty{0.005}{\radian} \approx \qty{17.2}{\arcminute}$.}
    \label{fig:hi_temp_map}
\end{figure} \\
Figures \ref{fig:dm_mass_map} and \ref{fig:hi_temp_map} show examples of a dark matter mass map and a HI brightness temperature map, respectively. The maps are for the frequency bin from 785 to \qty{790}{\mega\hertz} corresponding to redshifts from 0.798 to 0.809. For better visualisation, we present versions smoothed with a Gaussian kernel with $\sigma = \qty{0.005}{\radian} \approx \qty{17.2}{\arcminute}$. Unless a quantity is shown as a function of redshift, only the results for the 785 to \qty{790}{\mega\hertz} bin or the snapshot at $z = 0.8$ are shown in the following sections, as the results for the other frequency bins are similar.

\subsection{Mass Function}
\label{sec:Mass Function}
\texttt{PINOCCHIO} offers the option to output snapshots at different redshifts, i.e. dark matter halo catalogues of a single simulation box with mass, position, and velocity for each halo. It also directly provides the corresponding mass functions of these snapshots. We write out snapshots at $z = 0.8, 0.9, 1$. 
\begin{figure}
    \centering
    \includegraphics[width=0.9\textwidth]{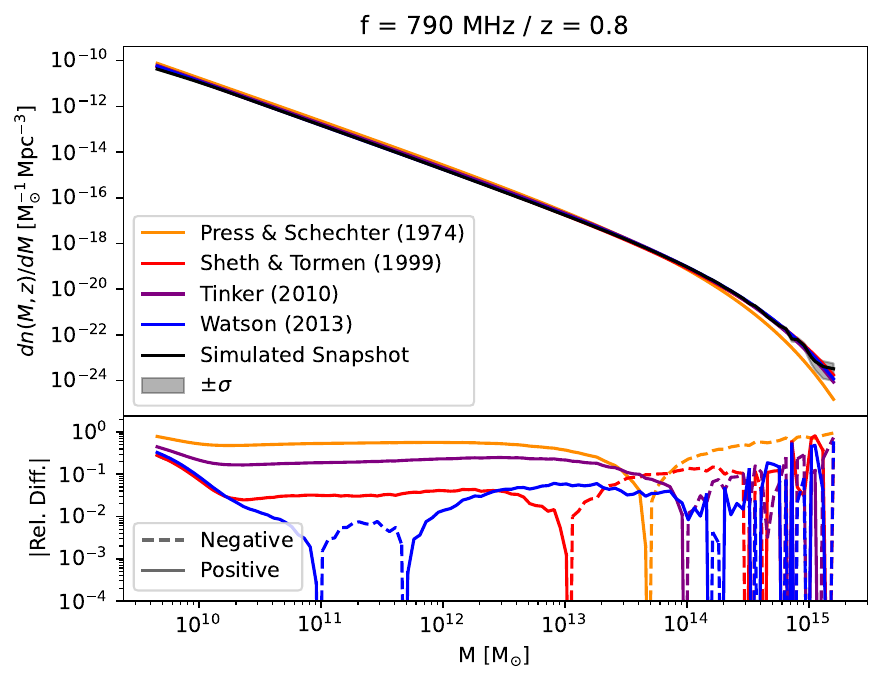}
    \caption{Comparison between mass functions implemented in \texttt{PyCosmo} (orange, red, purple, blue) \cite{Press_Schechter_1974, Sheth_Tormen_1999, Tinker_2010, Watson_2013} and the mass function measured on a snapshot of the simulation (black) at frequency \qty{790}{\mega\hertz} or redshift 0.8, respectively (upper panel). The grey shaded area indicates the Poisson error. The lower panel shows the absolute relative difference, with solid (dashed) lines indicating positive (negative) differences.}
    \label{fig:mass_function}
\end{figure} \\
Figure \ref{fig:mass_function} compares the measured mass function at $z = 0.8$ with the four options of the mass function implemented in \texttt{PyCosmo}. The lower part shows the absolute relative difference between the implementations and the measured function, distinguishing between positive (solid line) and negative differences (dashed line). We find that the mass functions of Watson and Sheth \& Tormen agree well with the simulation for intermediate halo masses, with agreement at the level of a few percent. At the low mass end, the simulation lacks halos. This is most likely due to our low choice of at least ten particles per halo. It is possible that the simulation may not be able to identify such small halos accurately. At the high mass end, the measured mass function is dominated by Poisson noise, as there are only a limited number of large mass halos in the snapshot. The grey shaded area indicates the Poisson noise of the snapshot. Overall, these observations are consistent with the mass function findings reported in \cite{PINOCCHIO_2017a}, where they find that \texttt{PINOCCHIO} reproduces the analytic mass function of Watson to within \qty{5}{\%}.

\subsection{Density}
\label{sec:Density}
\begin{figure}
    \centering
    \includegraphics[width=1.0\textwidth]{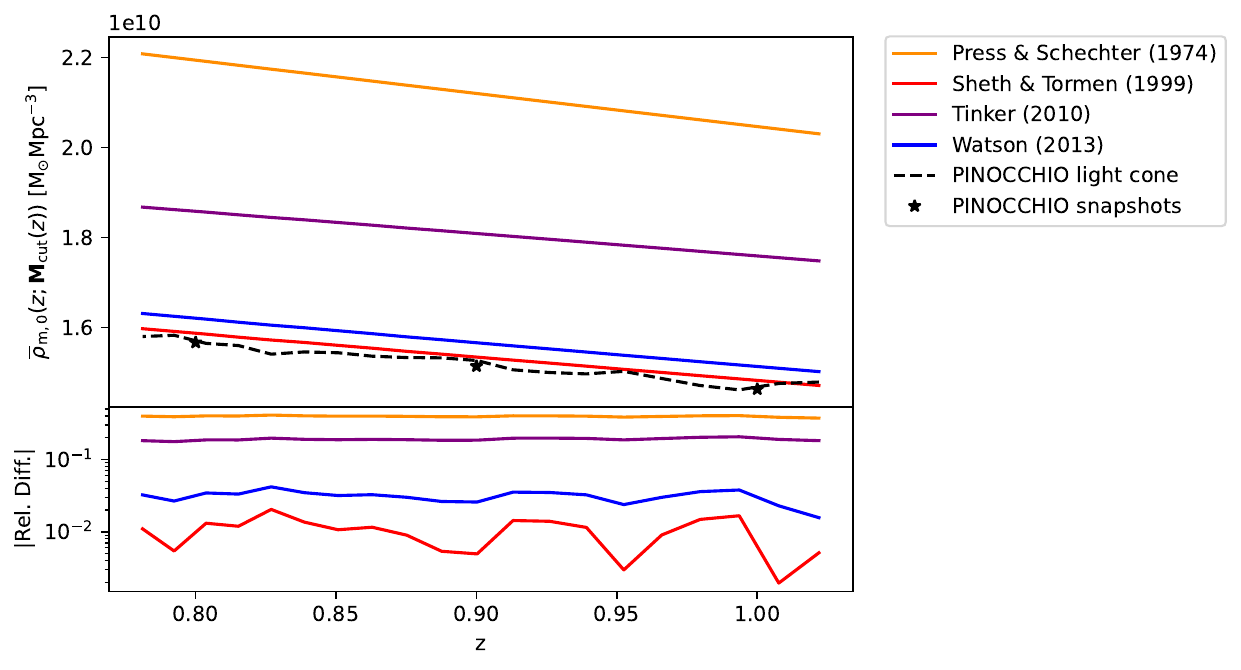}
    \caption{Comparison of the mean dark matter density over the simulated redshift range. The predicted densities for the different mass functions in \texttt{PyCosmo} (orange, red, purple, blue) \cite{Press_Schechter_1974, Sheth_Tormen_1999, Tinker_2010, Watson_2013} are compared with the density computed from the snapshots of the simulation at $z = 0.8, 0.9, 1$ (black stars), and with the density computed from the maps we created (black dashed line). In the lower panel, we show the absolute relative difference between the different lines to the density measured on the light cone.}
    \label{fig:rho_m}
\end{figure}

\begin{figure}
    \centering
    \includegraphics[width=1.0\textwidth]{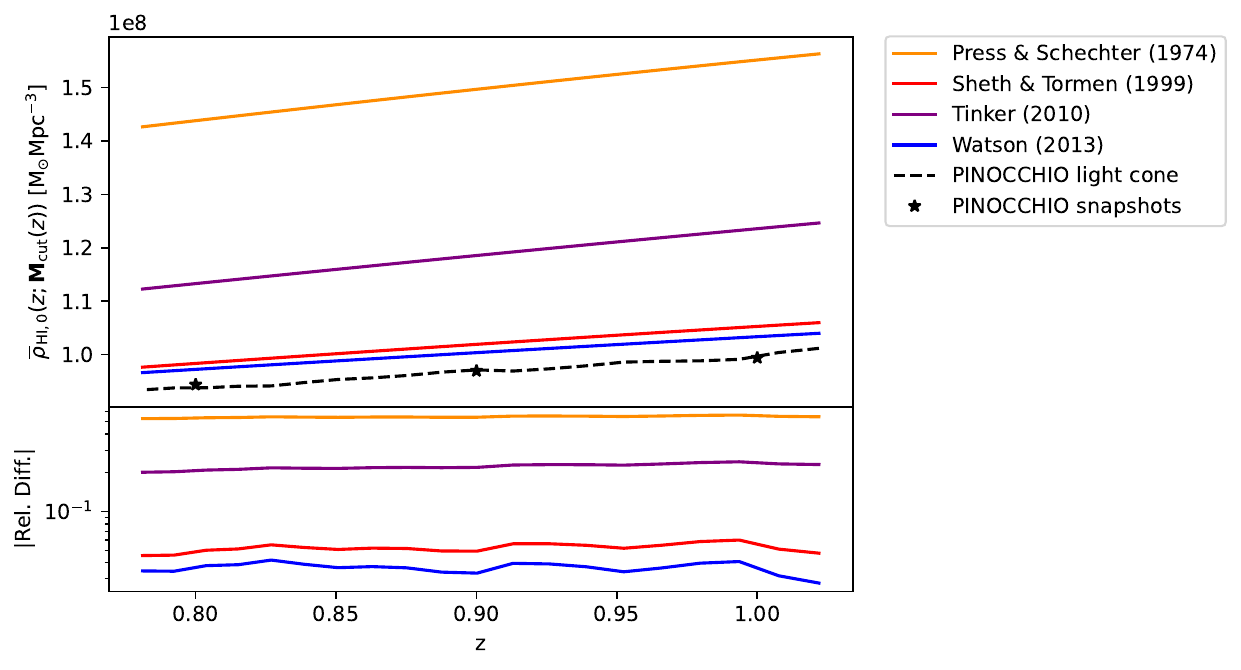}
    \caption{Same comparison as in figure \ref{fig:rho_m}, but for the HI mass density, equation (\ref{eq:HI rhobar}), instead of dark matter density.}
    \label{fig:rho_hi}
\end{figure}

In a next step, the mass functions weighted by the mass can be integrated over the simulated mass range, as described in equations (\ref{eq:matter rhobar}) and (\ref{eq:HI rhobar}), to predict the mean dark matter and HI mass densities for the different frequency bins. Figures \ref{fig:rho_m} and \ref{fig:rho_hi} compare the mass densities calculated with the different mass functions in \texttt{PyCosmo} to the simulated densities. We check the snapshots and the maps that are generated based on the past light cone. The density in a snapshot is calculated by integrating the measured mass function. The density in a map is calculated by summing the total mass in a map and dividing it by the volume projected onto the map.
 \\
As mentioned in section \ref{sec:Ingredients}, the densities considered here only capture the hierarchical growth of the halos, but neglect the expansion of the universe. This explains the increasing dark matter density with decreasing redshift. In the case of HI, it must also be taken into account that in the HI model used here (see equation (\ref{eq:MHI-M relation})) the HI fraction in low-mass halos, with virial velocities below the cut-off velocity $v_\text{c,0}$, decreases as time evolves. Therefore, the HI mass density decreases with decreasing redshift. \\
There is good agreement between the densities calculated from the snapshots and the maps, i.e. the past light cone. The lower panels show the relative difference between the \texttt{PyCosmo} predictions and the densities in the maps. As expected from the comparisons in figure \ref{fig:mass_function}, the mass functions from Watson and Sheth \& Tormen also give the best agreement for the densities compared to the simulation. For dark matter, the Sheth \& Tormen mass function agrees slightly better with a 0.2 -- \qty{2}{\%} difference than the Watson mass function with 1.5 -- \qty{4.2}{\%}. Similar results are found in \cite{PINOCCHIO_2017a}, where the number density of halos more massive than \qty{9.9e12}{M_{\odot}} and \qty{4.9e13}{M_{\odot}} in the past light cone is compared to the number density in snapshots and a prediction based on the Watson mass function. They find very good agreement between the snapshots and the light cone and again within \qty{5}{\%} when compared to Watson. For HI, the Watson mass function with a relative difference of 2.8 -- \qty{4.2}{\%} achieves a better agreement than the Sheth \& Tormen mass function, for which we find an agreement of 4.5 -- \qty{6}{\%}. This is due to the different mass weighting when considering dark matter halos or HI halos.

\subsection{Power Spectrum}
\label{sec:Power Spectrum}
\begin{figure}
    \centering
    \includegraphics[width=0.9\textwidth]{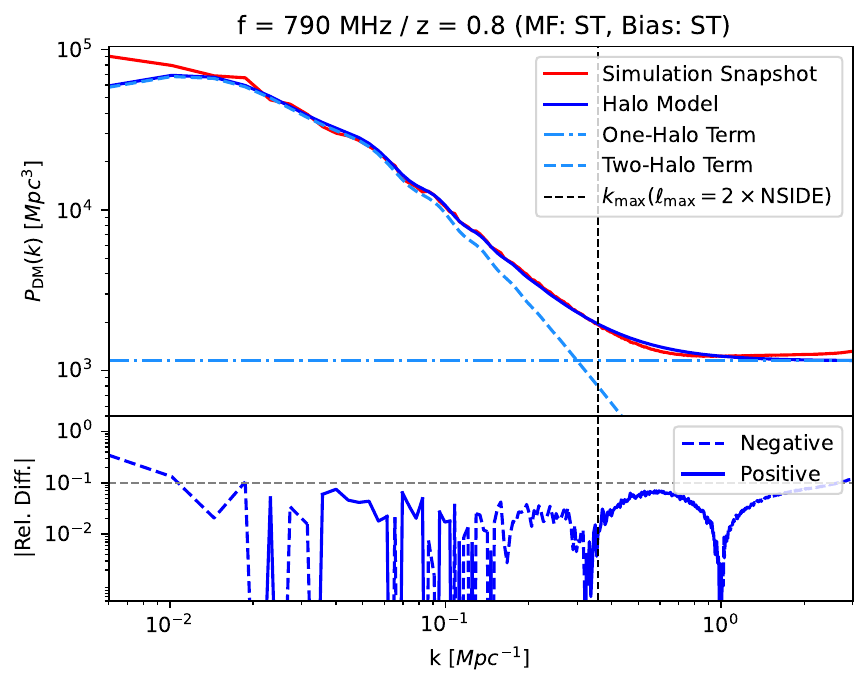}
    \caption{Dark matter power spectrum of a simulation snapshot (red) at \qty{790}{\mega\hertz}, or redshift 0.8, compared to the halo model prediction (blue). In addition, the contributions of the one-halo term (dashed-dotted light blue line) and two-halo term (dashed light blue line) are shown. We assume the mass function and linear halo bias of Sheth \& Tormen \cite{Sheth_Tormen_1999} in the halo model calculations. The vertical dashed black line indicates $k_\mathrm{max}$, which corresponds to the maximum multipole $\ell_\mathrm{max}$ that is considered in the angular power spectrum in section \ref{sec:Angular Power Spectrum}. In the bottom panel, we show the absolute relative difference between the halo model and the simulation. The horizontal dashed grey line indicates the \qty{10}{\%} relative difference.}
    \label{fig:pk_dm}
\end{figure}

\begin{figure}
    \centering
    \includegraphics[width=0.9\textwidth]{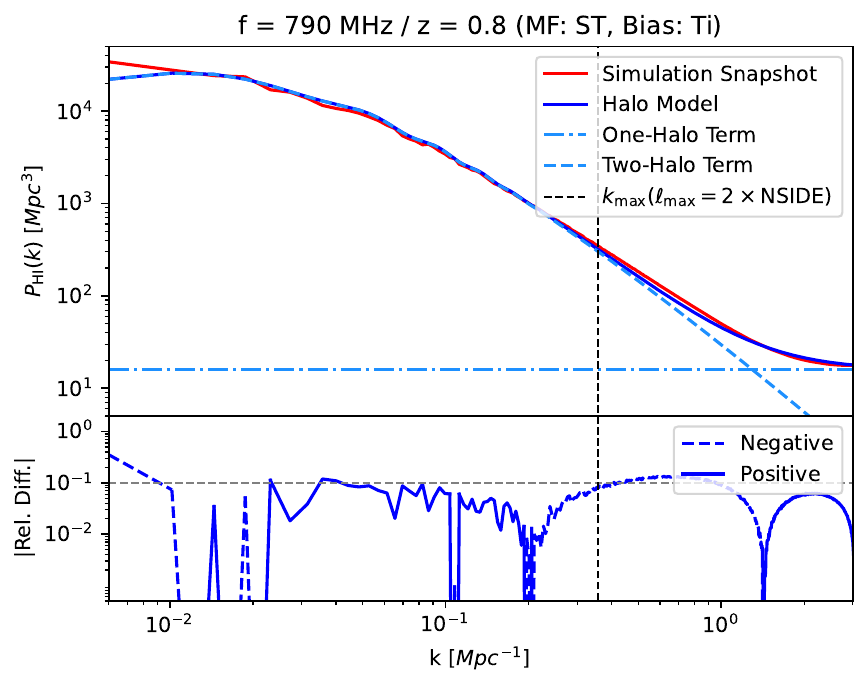}
    \caption{Same comparison as in figure \ref{fig:pk_dm}, but for the HI power spectrum instead of the dark matter power spectrum. Here, we assume the Sheth \& Tormen mass function \cite{Sheth_Tormen_1999} and the linear halo bias from Tinker \cite{Tinker_2010}.}
    \label{fig:pk_hi}
\end{figure}

In figures \ref{fig:pk_dm} and \ref{fig:pk_hi}, the three-dimensional power spectra for dark matter and HI calculated from the snapshot are compared with the halo model predictions. To compute the three-dimensional power spectra of the snapshot the Python libraries from \texttt{Pylians} \cite{Pylians} have been used. The snapshot provided by \texttt{PINOCCHIO} can be used directly to compute the dark matter power spectrum. For HI, however, the dark matter halo mass must first be converted into HI mass using equation (\ref{eq:MHI-M relation}) again. Both figures explicitly show the contributions of the one- and two-halo terms for the theoretical predictions with the halo model. The black dashed line indicates $k_\mathrm{max} = (\ell_\mathrm{max} + 1/2) / r(\chi(z))$, \footnote{Note that this correspondence between \textit{k} and $\ell$ is taken from the argument of the power spectrum in the expression of the Limber approximation.} which corresponds to the maximum multipole moment for which we calculate the angular power spectrum later in section \ref{sec:Angular Power Spectrum}. 
 \\
In the previous two sections, we have seen that the mass functions of Watson or Sheth \& Tormen agree best with the simulation. When the measured mass function of the snapshot is used to calculate the power spectrum and the linear halo bias is varied, it is found that for the three snapshots at different redshifts, the bias from the Sheth \& Tormen mass function best matches the power spectrum of the snapshot for dark matter and the bias from the Tinker mass function for HI. Looking at all the combinations of mass functions and linear halo biases in \texttt{PyCosmo}, it is found that for dark matter, the combination of the Sheth \& Tormen mass function and linear halo bias best fits the power spectrum of the snapshot. For HI, it is the combination of the Sheth \& Tormen mass function and the linear halo bias from the Tinker mass function. \footnote{For HI, the combination of the bias derived from the Tinker mass function with the mass functions from either Watson or Tinker also agrees well with the measured power spectrum.} \\
As expected, the two-halo term is dominant on large scales, i.e. for small \textit{k}'s, describing the correlation between different halos, while the one-halo term is dominant on small scales, i.e. for large \textit{k}'s, describing the correlation within a single halo. The latter contributes as an overall constant shot-noise because the halos are assumed to be point sources with no radial profile, as previously described in sections \ref{sec:DM Power Spectrum} and \ref{sec:HI Power Spectrum}, in agreement with the simulations. \\
We again plot the absolute relative difference between the halo model prediction and the spectrum of the snapshot and find good agreement within a few percent over a broad range of scales for both dark matter and HI. For dark matter this level of agreement is consistent with similar results reported in \cite{PINOCCHIO_2017a} where the power spectrum of \texttt{PINOCCHIO} is compared to an \textit{N}-body simulation within the range of $\qty{2e-2}{Mpc^{-1}} < k < \qty{3e-1}{Mpc^{-1}}$. The agreement we find here is particularly remarkable given the simplicity of the halo model, especially for HI, since the mass functions and linear halo biases were originally calibrated on dark matter halo simulations. That could potentially explain why the agreement is slightly worse for HI compared to dark matter. The discrepancy between the simulation and the halo model observed for $k < 10^{-2} \, \text{Mpc}^{-1}$ arises from the limited box size of our simulation, i.e. \qty{1}{h^{-1} Gpc}. This box size corresponds to $k \approx \qty{4e-3}{Mpc^{-1}}$, and as a result, modes on scales comparable to that are sampled less often which results in weak statistical power. \\
We are aware of the optimized variant of the halo model presented in \cite{Mead_2015, Mead_2016}, where physically motivated free parameters have been introduced and determined by fitting to simulated power spectra. While these parameters are well-motivated for dark matter, their validity for HI is not guaranteed. To enable a clear comparison between the dark matter and HI results in this work, we refrained from introducing the adapted dark matter halo model. However, the model described in \cite{Mead_2015, Mead_2016} is implemented in \texttt{PyCosmo}, though not made use of in this work.

\subsection{Angular Power Spectrum}
\label{sec:Angular Power Spectrum}
\begin{figure}
    \centering
    \includegraphics[width=1.0\textwidth]{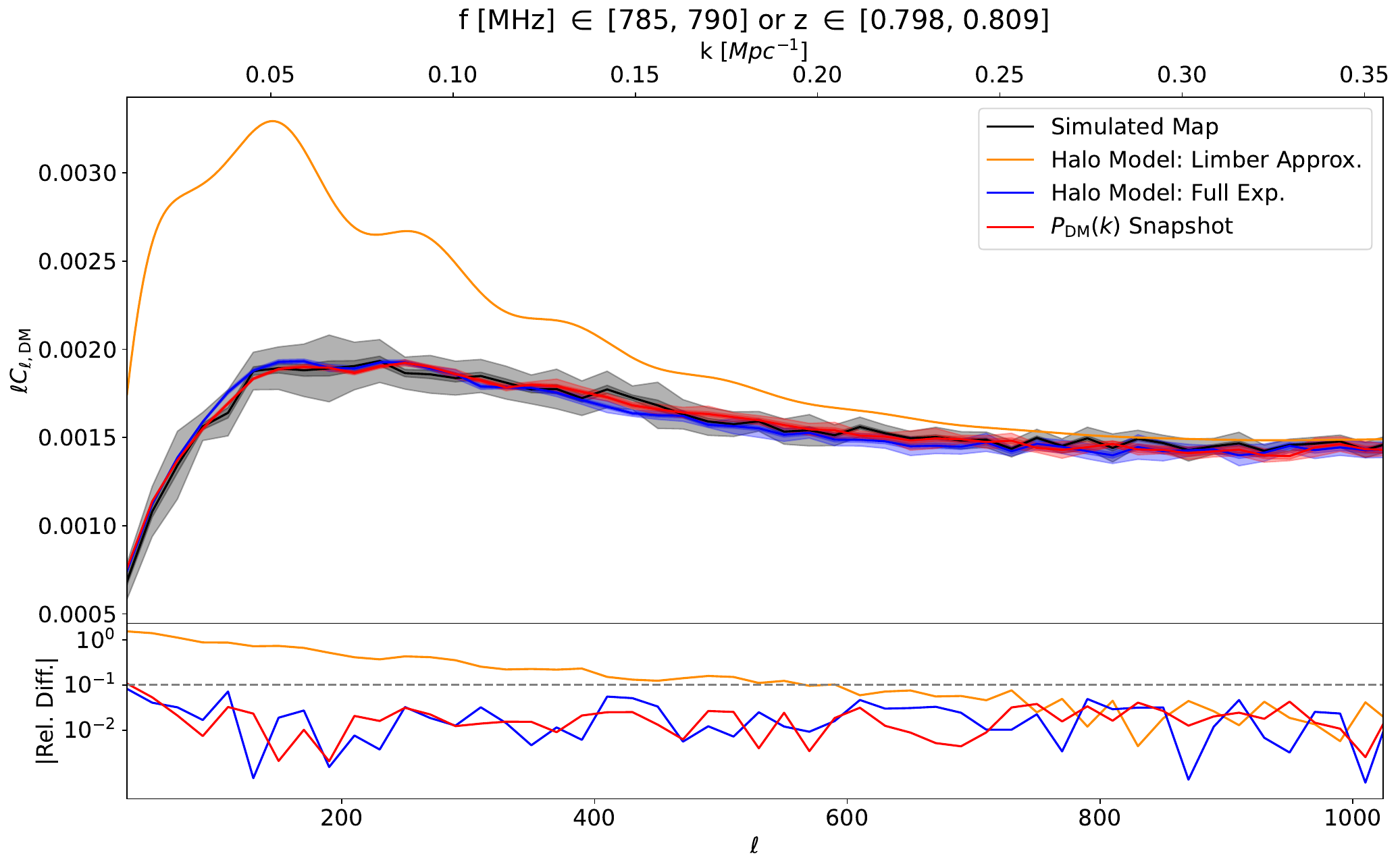}
    \caption{Angular Power Spectrum for the dark matter overdensity map for the frequency bin from 785 to \qty{790}{\mega\hertz} (black). We show the mean $C_{\ell}$'s averaged over 20 adjacent $\ell$'s. The dark grey band indicates the error on the mean and the light grey band the standard deviation. The spectrum of the map is compared with the halo model predictions: The orange line shows the Limber approximation. The blue and red lines show the spectra resulting from the integration without the approximation for the $P(k)$ based on the halo model or the snapshot, respectively (see equation (\ref{eq:Cl_dm})). The three-dimensional integration is performed with a Monte-Carlo integration. The blue and red shaded bands indicate the error of the integration. The bottom panel shows the absolute relative difference and the grey dashed line indicates \qty{10}{\%}.}
    \label{fig:cl_dm}
\end{figure}

\begin{figure}
    \centering
    \includegraphics[width=1.0\textwidth]{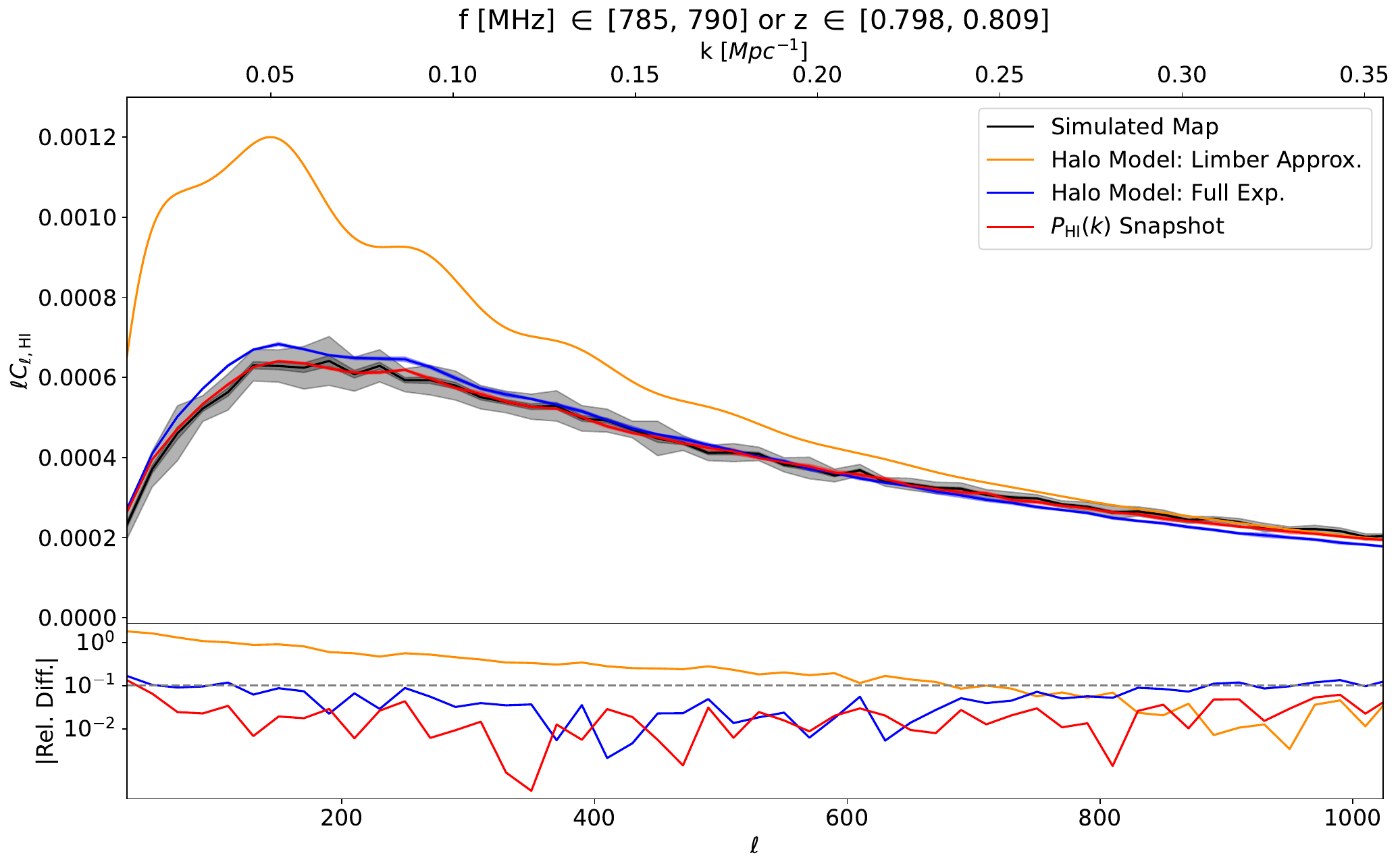}
    \caption{Same comparison as in figure \ref{fig:cl_dm} but for the HI angular power spectrum.}
    \label{fig:cl_hi}
\end{figure}

Finally, the angular power spectra of the maps are calculated. For comparison with the halo model expressions described in sections \ref{sec:DM Angular Power Spectrum} and \ref{sec:HI Angular Power Spectrum}, we calculate overdensity maps by subtracting and dividing by the mean pixel mass for the dark matter maps and the mean pixel brightness temperature for the HI maps, respectively. The \textit{anafast} function of \texttt{HEALPix} is then used to calculate the angular power spectrum. As recommended in the documentation \footnote{\url{https://healpix.sourceforge.io/html/fac_anafast.htm}.}, the angular power spectrum is calculated up to a multipole moment of $\ell_\mathrm{max} = 2 \times NSIDE = 1024$ for $NSIDE = 512$. For a clean comparison between the halo model and the maps, one has to correct the theoretical predictions for the partial sky coverage and the pixel window function. \footnote{\url{https://healpix.sourceforge.io/html/intro_Pixel_window_functions.htm}} For the former, it was verified that dividing the angular power spectrum expression from the halo model by the sky coverage is sufficient in the multipole range of interest, with no further deconvolution of the sky mask required. For the latter, it is essential to subtract the Poisson shot-noise of the simulation before applying the correction. The pixel window function represents the smoothing of the signal from individual sources across the pixel area. However, the Poisson noise arises due to the discrete nature of the halos and is therefore independent of pixelization. Applying the pixel window function to the Poisson noise would result in an overestimation of the signal measured from the maps. By adapting the expression from \cite{Nicola_2020} for mass overdensity maps, the Poisson contribution to the angular power spectrum can be calculated as:
\begin{equation}
    C_{\ell}^{SN} = \Omega_\text{pix} \frac{\big<m^2\big>}{\overline{N} \, \overline{m}^2},
    \label{eq:Poisson Noise}
\end{equation}
where $\Omega_\text{pix}$ is the pixel area in steradians, $\big<m^2\big>$ the average of the squared halo masses in the map, $\overline{N}$ the average number of halos per pixel, and $\overline{m}$ the mean halo mass in the map. After applying the pixel window function correction, the Poisson noise is reintroduced to the signal. Figures \ref{fig:cl_dm} and \ref{fig:cl_hi} show the angular power spectrum for dark matter and HI calculated from the maps (black). For better visualisation, the $C_{\ell}$'s are averaged over 20 neighbouring $\ell$'s. The dark grey band shows the error of the mean value and the light grey band the standard deviation calculated on the bins. 
 \\
The $C_{\ell}$'s of the maps are compared to the halo model. As for the power spectra in the last section, the combinations of the Sheth \& Tormen mass function and linear halo bias for the dark matter maps and the Tinker bias for HI have been used. In both cases, it is apparent that the Limber approximation (orange), i.e. equations (\ref{eq:Cl_dm_limber}) and (\ref{eq:Cl_HI_limber}), strongly overestimates the spectra due to the thinness of the maps for which the approximation diverges. The full expressions for the spectra (blue), equations (\ref{eq:Cl_dm}) and (\ref{eq:Cl_HI}), agree much better. As for the power spectra, we again find very good agreement below \qty{10}{\%} over all shown scales. For the angular power spectra shown, we integrate over the \textit{k}-range that is given in figures \ref{fig:pk_dm} and \ref{fig:pk_hi}, i.e. $\sim 6 \times 10^{-3}$ -- \qty{3}{Mpc^{-1}}. It has been further checked that this \textit{k}-range is sufficiently large for the integration to converge. The shaded areas give the errors of the Monte-Carlo integrations. \\
We also calculate the $C_{\ell}$'s with equations (\ref{eq:Cl_dm}) and (\ref{eq:Cl_HI}) by using the measured $P(k)$ of the snapshots as discussed in the last section instead of the halo model. To facilitate the comparison between the power spectrum and the angular power spectrum, we highlight $k_\text{max}$ in figure \ref{fig:pk_dm} and \ref{fig:pk_hi}, which corresponds to the maximum multipole moment shown in figure \ref{fig:cl_dm} and \ref{fig:cl_hi}. \footnote{The same relation between \textit{k} and $\ell$, given in section \ref{sec:Power Spectrum}, is used to give the corresponding \textit{k}-values on the upper horizontal axis for the angular power spectra.} As already with the power spectra for dark matter, the angular power spectra match very well too. For HI, the agreement with the angular power spectra of the maps for $\ell < 350$ and $\ell > 700$ gets better when the snapshot power spectrum is used. Therefore, the differences seen in the $P_\text{HI}(k)$ transfer to the angular power spectrum and explain the discrepancies that we observe there. \\
As discussed in section \ref{sec:Simulation of Dark Matter Halos and HI Halos}, scales comparable to or larger than the simulated box size are affected by box replications. The box size corresponds to $\ell \approx 10$, but as highlighted in section \ref{sec:Power Spectrum}, already scales $k < 10^{-2} \, \text{Mpc}^{-1}$ or $\ell < 30$ are impacted by the limited box size. While a disagreement becomes apparent at these scales, the plots do not emphasize this region.

\section{Conclusions}
\label{sec:Conclusions}
Post-reionization \qty{21}{\cm} intensity mapping is a promising, powerful and efficient cosmological probe. By inherently encoding the position along the line of sight within the redshifted signal from unresolved neutral hydrogen, this method eliminates the need for spectroscopic follow-up of individual galaxies to determine their redshifts. This holds the potential to cover large volumes in a short time, enabling highly accurate constraints on cosmological parameters. The increasing recognition of the importance of this probe is evident in the growing number of radio interferometric projects that have been realised or are being planned. Simulations of the \qty{21}{\cm} signal are invaluable for testing observation modeling and analysis pipelines, as well as for conducting forecast studies of future surveys and parameter constraints in the context of simulation-based inference. \\
In this paper, we developed a method to generate realistic post-reionization \qty{21}{\cm} emission maps. Utilizing the approximate but efficient \texttt{PINOCCHIO} code, we simulated the past light cone of dark matter halos and populated them with HI according to an HI-halo mass relation. This approach enabled the creation of large-volume simulations with high halo mass resolution, essential for \qty{21}{\cm} intensity mapping. We focused on simulating a past light cone specified for the HIRAX radio interferometer, i.e. covering declinations between \qty{-15}{\degree} and \qty{-35}{\degree} in the frequency range from 700 to \qty{800}{\mega\hertz}, i.e. a redshift range of $0.77 < z < 1.03$. We ran \texttt{PINOCCHIO} for a box size of \qty{1}{h^{-1} Gpc} including $6700^3$ simulation particles. It needs 40 box replications which are stacked together with harmonic boundary conditions to cover the entire light cone. We simulate dark matter halos down to masses of \qty{4.3e9}{M_{\odot}}, achieving an HI loss of only 2 -- \qty{3}{\%} within the simulated redshift range. Subsequently, we generated dark matter and HI signal maps by dividing the light cone into 20 equally spaced frequency channels with a width of \qty{5}{\mega\hertz} to account for the radial resolution of the signal. \\
To validate the quality of our simulation and the generated maps, we implemented a dark matter halo model and an adapted HI halo model in \texttt{PyCosmo}. Based on the basic assumption that all dark matter resides in halos of any size and assuming a halo mass function, a linear halo bias, and a halo profile, the models provide predictions for the mass densities, the non-linear power spectra, and consequently the angular power spectra. The implementations are flexible enough to account for the simulated mass resolution, allowing direct comparison with the simulations. We provide four different halo mass functions \cite{Press_Schechter_1974, Sheth_Tormen_1999, Tinker_2010, Watson_2013} and four linear halo biases \cite{Sheth_Tormen_1999, Tinker_2010, Sheth_Mo_Tormen_2001, Mo_White_1996} in \texttt{PyCosmo}. We offer NFW profiles for dark matter and exponential profiles for HI halos. However, the simulation treats halos as point sources without substructure. Therefore, we compared the models to the simulation using Dirac delta profiles. Currently, \texttt{PyCosmo} only includes the angular power spectrum assuming the Limber approximation. We found that this approximation does not apply very well to thin maps. Hence, we considered the full calculation without the approximation. But because it consists of an expensive three-dimensional integration including spherical Bessel functions, we perform it outside of \texttt{PyCosmo} with a Monte-Carlo integration method. \\
Our findings show that the models and the simulation agree on the mass function and the mass densities within a few percent. We also find good agreement below \qty{10}{\%} for the power spectrum and angular power spectrum of dark matter and HI. This agreement, consistent with the expectations of a simple halo model and the approximate \texttt{PINOCCHIO} code, gives us confidence in the quality and understanding of the simulated signal. \\
To avoid complicating the analysis, we have made the comparison between the simulation and theory in real space, considering the actual redshift of the halos. However, redshift measurements are always performed in redshift space, introducing redshift space distortions due to the peculiar velocities of the halos. Fortunately, \texttt{PINOCCHIO} provides the velocities and the observed redshifts in the halo catalogue, making it straightforward to include them from the simulation perspective. On the theory side, distortion terms must be included in the power spectrum, as for example described in \cite{Padmanabhan_2023}. Dedicated work on these implementations has already started and will continue henceforth. \\
In future work, we plan to apply the maps to the instrument simulation and analysis pipeline of the HIRAX instrument, which is based on the \textit{m-mode} formalism described in \cite{Shaw_2014, Shaw_2015}. Our goal is to incorporate instrumental noise along with galactic and extra-galactic foregrounds that contaminate the cosmological signal. The latter poses probably the foremost challenge for \qty{21}{\cm} cosmology. We aim to test the pipeline's ability to remove the foregrounds and retrieve the underlying \qty{21}{\cm} signal. Additionally, we will investigate the pipeline's sensitivity to variations in cosmological parameters and astrophysics, such as alternative HI-halo mass relations, within our simulations. \\
Cross-correlating the \qty{21}{\cm} intensity mapping signal with other cosmological probes, as demonstrated in previous studies \cite{MeerKLASS_2024, CHIME_2023, Cunnington_2022, Wolz_2021, Masui_2013, Chang_2010}, allow for detection of the HI signal with reduced systematics. Our approach facilitates future cross-correlation studies between \qty{21}{\cm} intensity mapping and other probes by providing consistent halo catalogues that can be used to model other observables. One such probe could be CMB lensing. In \cite{Moodley_2023}, they illustrate that HI intensity mapping suffers from long-wavelength line-of-sight mode loss due to galactic foreground subtraction, which significantly reduces the constraining power of the cross-correlation power spectrum with CMB lensing. Nonetheless, a proposed cross-bispectrum estimator, which cross-correlates the HI power spectrum with the CMB lensing field, enhances constraints on cosmological parameters. Forecasts combining hypothetical HIRAX HI intensity mapping measurements with CMB lensing measurements from ground based CMB experiments show that this estimator can yield competitive constraints on cosmological parameters, particularly on the dark energy equation of state parameters, compared to next-generation galaxy redshift surveys. Reinforcing these results with realistic simulations would confirm that, despite the significant challenge posed by foregrounds, \qty{21}{\cm} intensity mapping remains a promising cosmological probe. \\
\texttt{PyCosmo 2.2.0} with the halo model extension is available at \url{https://cosmology.ethz.ch/research/software-lab/PyCosmo.html} and can be installed from the Python Package Index at \url{https://pypi.org}. The simulation dataset is available at \url{https://cosmology.ethz.ch/research/software-lab/cosmological-neutral-hydrogen-simulation.html}. The dataset includes \texttt{PINOCCHIO}'s dark matter halo catalogues of the snapshots at $z = 0.8, 0.9, 1$ and the past light cone saved as text files as well as the dark matter mass and HI brightness temperature maps for the 20 frequency bins stored as HDF5 files.

\acknowledgments
We thank Marta Spinelli, Sara Aliqolizadehsafari, Jennifer Studer, and Luis Machado Poletti Valle for helpful discussions and ongoing work to apply our simulations to instruments other than HIRAX and therefore help spreading awareness. We thank Hamsa Padmanabhan for insightful discussions on the halo model. We thank Uwe Schmitt for his continuous software support throughout the project, especially with respect to \texttt{PyCosmo}. We thank Pierluigi Monaco for his support in the use of \texttt{PINOCCHIO}. We thank Andrina Nicola for her great help in interpreting the angular power spectrum. \\
We acknowledge access to Piz Daint at the Swiss National Supercomputing Centre, Switzerland, under the SKA’s share with the project IDs go06, sk05, and sk19. \\
This work was supported in part by grant 200021\_192243 from the Swiss National Science Foundation, and by SERI as part of the SKACH consortium. \\
This work made use of the \texttt{numpy} \cite{numpy}, \texttt{scipy} \cite{scipy}, \texttt{matplotlib} \cite{matplotlib}, \texttt{h5py} \cite{h5py}, and \texttt{Pylians} \cite{Pylians} software packages. Some of the results in this paper have been derived using the \texttt{healpy} \cite{healpy} and \texttt{HEALPix} \cite{HEALPix} package.

\appendix
\section{Poisson Noise Contribution to the Angular Power Spectrum}
\label{sec:Poisson Noise Contribution to the Angular Power Spectrum}
\begin{figure}
    \centering
    \includegraphics[width=1\linewidth]{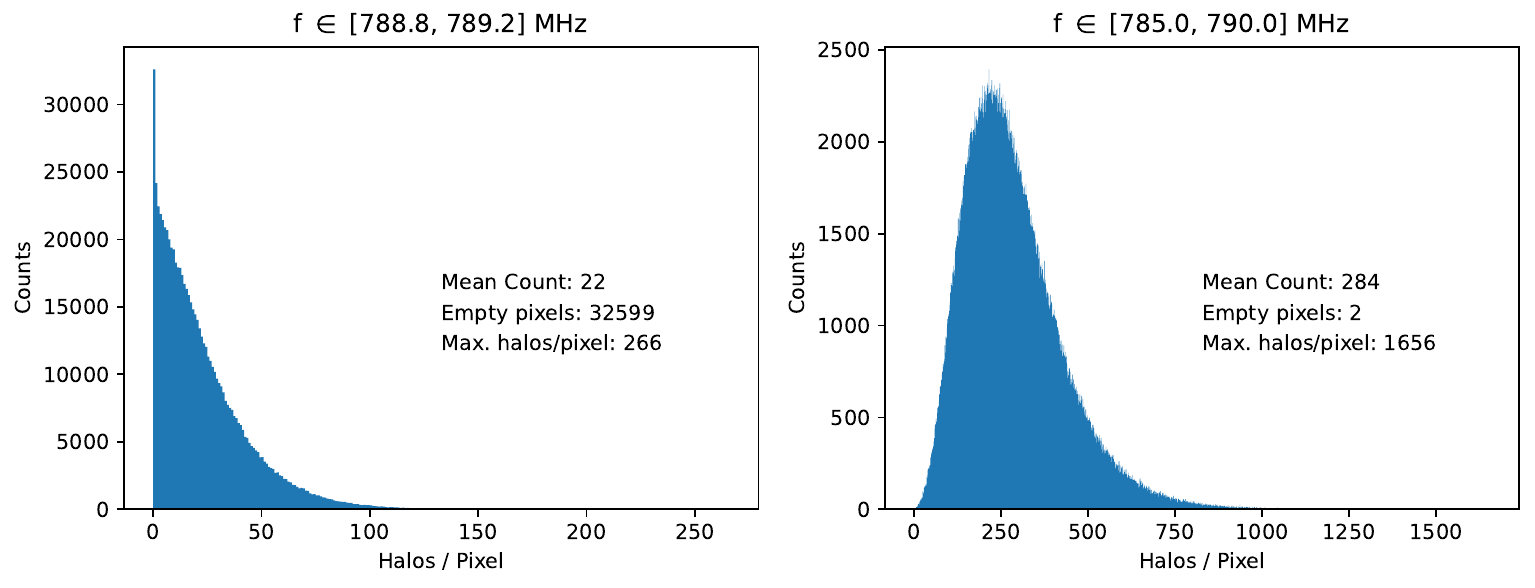}
    \caption{Number of halos per pixel for maps with bandwidths of \qty{0.4}{\mega \hertz} (left) and \qty{5}{\mega \hertz} (right).}
    \label{fig:halo_counts_comparison}
\end{figure}
Figure \ref{fig:halo_counts_comparison} shows the halo counts per pixel for a bandwidth of \qty{0.4}{\mega \hertz} (left) and \qty{5}{\mega \hertz} (right). The former shows a more Poisson-like distribution with a much lower mean count and many more empty pixels compared to the latter, which shows a more lognormal distribution.
\begin{figure}
    \centering
    \includegraphics[width=1.0\linewidth]{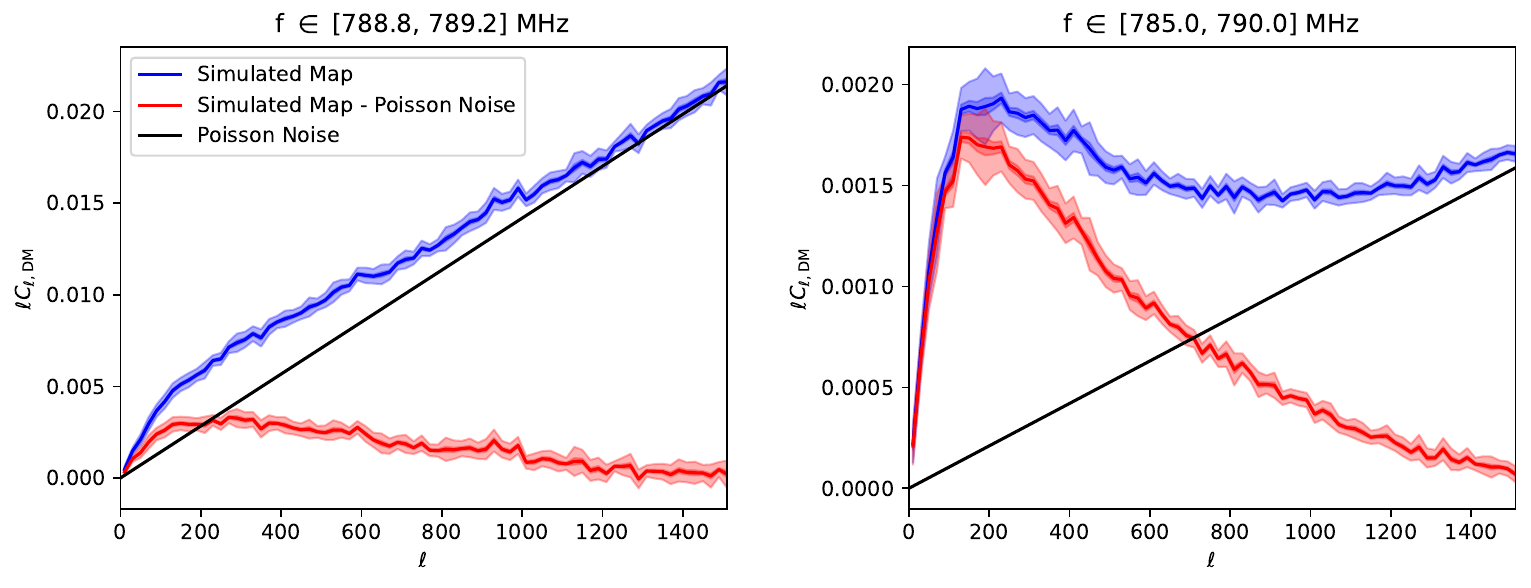}
    \caption{Poisson noise contribution to the angular power spectrum for dark matter maps with bandwidths of \qty{0.4}{\mega \hertz} (left) and \qty{5}{\mega \hertz} (right). For the simulated maps, we show the mean $C_{\ell}$ values averaged over 20 adjacent $\ell$'s. The dark blue/red band represents the error on the mean, while the light blue/red band indicates the standard deviation.}
    \label{fig:dm_shot_noise}
\end{figure}
\begin{figure}
    \centering
    \includegraphics[width=1.0\linewidth]{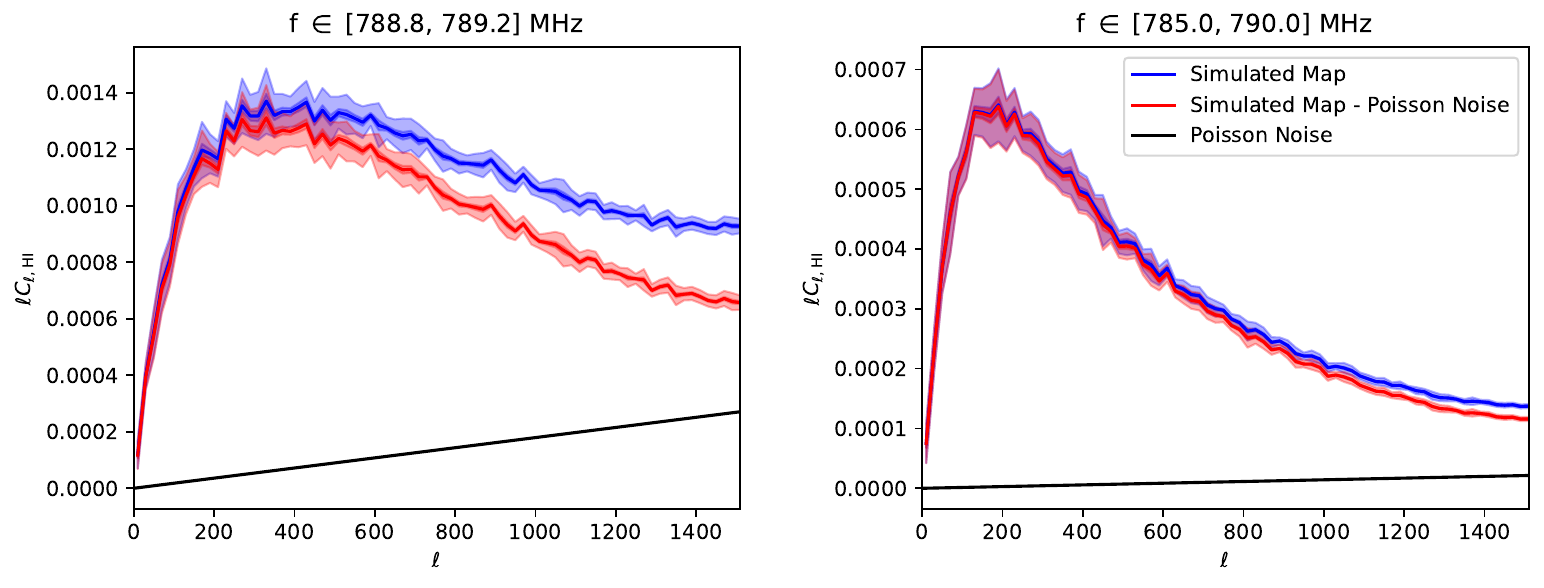}
    \caption{Poisson noise contribution to the angular power spectrum for HI brightness temperature maps with bandwidths of \qty{0.4}{\mega \hertz} (left) and \qty{5}{\mega \hertz} (right). As in figure \ref{fig:dm_shot_noise}, the dark blue/red band represents the error on the mean $C_{\ell}$ values averaged over 20 adjacent $\ell$'s and the light blue/red band the standard deviation.}
    \label{fig:hi_shot_noise}
\end{figure}
Figures \ref{fig:dm_shot_noise} and \ref{fig:hi_shot_noise} show the contributions of Poisson noise to the angular power spectra for bandwidths of \qty{0.4}{\mega \hertz} and \qty{5}{\mega \hertz} for dark matter and HI brightness temperature maps, respectively. For maps weighted by the dark matter halo masses, it can be clearly seen that for a bandwidth of \qty{0.4}{\mega \hertz} the Poisson noise (black) dominates over the cosmological signal (red) for $\ell > 200$, while this happens only for $\ell > 700$ for a bandwidth of \qty{5}{\mega \hertz}. \\
For HI, the contribution of Poisson noise is also larger for thinner maps, but overall less significant than for dark matter. This is due to the different mass weighting of the halos: Since the logarithmic slope in equation (\ref{eq:MHI-M relation}) is $\beta = -0.58$, $M_\text{HI}(M,z) \propto M^{0.42}$. This means that when weighting the halo mass function by the HI mass, the smaller and more abundant halos are weighted more than the larger and rarer halos. The Poisson noise is inversely proportional to the number of halos (see equation (\ref{eq:Poisson Noise})) and hence small halos have a lower Poisson noise than large halos. This results in a lower Poisson noise for HI maps than dark matter maps.

\end{document}